\documentclass[epjmod,nopacs,a4paper,10pt]{svjourmod}
\usepackage[english]{babel}
\usepackage{amsmath}
\usepackage{amssymb}
\usepackage{graphicx}
\usepackage{units}
\usepackage{subfigure}
\usepackage{cite}

\newcommand{\hb} {HERA-B\ }
\newcommand{\kss}{$K^{*0}$}

\newcommand{\kssb}{$\overline{K}^{\ast 0}$}
\newcommand{\kssto}{$K^{\ast 0}\to K^+\pi^-\ $}
\newcommand{\ksstop}{$K^{\ast 0}\to K^+\pi^-$}
\newcommand{\kssbto}{$\overline{K}^{\ast 0}\to K^-\pi^+\ $}

\begin{document}

\title{\kss\ and $\phi$ Meson Production in Proton-Nucleus
  Interactions at $\sqrt{s} = \unit[41.6]{GeV}$}

\author{
I.~Abt\inst{23}\and
M.~Adams\inst{10}\and
M.~Agari\inst{13}\and
H.~Albrecht\inst{12}\and
A.~Aleksandrov\inst{29}\and
V.~Amaral\inst{8}\and
A.~Amorim\inst{8}\and
S.~J.~Aplin\inst{12}\and
V.~Aushev\inst{16}\and
Y.~Bagaturia\inst{12\and36}
% ...
}

\mail{Christopher.van.Eldik@mpi-hd.mpg.de}

\institute{
$^{1}${\it NIKHEF, 1009 DB Amsterdam, The Netherlands~$^{a}$} \\
$^{2}${\it Department ECM, Faculty of Physics, University of Barcelona, E-08028 Barcelona, Spain~$^{b}$} \\
$^{3}${\it Institute for High Energy Physics, Beijing 100039, P.R. China} \\
$^{4}${\it Institute of Engineering Physics, Tsinghua University, Beijing 100084, P.R. China} \\
$^{5}${\it Institut f\"ur Physik, Humboldt-Universit\"at zu Berlin, D-12489 Berlin, Germany~$^{c,d}$} \\
$^{6}${\it Dipartimento di Fisica dell' Universit\`{a} di Bologna and INFN Sezione di Bologna, I-40126 Bologna, Italy} \\
$^{7}${\it Department of Physics, University of Cincinnati, Cincinnati, Ohio 45221, USA~$^{e}$} \\
$^{8}${\it LIP Coimbra, P-3004-516 Coimbra,  Portugal~$^{f}$} \\
$^{9}${\it Niels Bohr Institutet, DK 2100 Copenhagen, Denmark~$^{g}$} \\
$^{10}${\it Institut f\"ur Physik, Universit\"at Dortmund, D-44221 Dortmund, Germany~$^{d}$} \\
$^{11}${\it Joint Institute for Nuclear Research Dubna, 141980 Dubna, Moscow region, Russia} \\
$^{12}${\it DESY, D-22603 Hamburg, Germany} \\
$^{13}${\it Max-Planck-Institut f\"ur Kernphysik, D-69117 Heidelberg, Germany~$^{d}$} \\
$^{14}${\it Physikalisches Institut, Universit\"at Heidelberg, D-69120 Heidelberg, Germany~$^{d}$} \\
$^{15}${\it Department of Physics, University of Houston, Houston, TX 77204, USA~$^{e}$} \\
$^{16}${\it Institute for Nuclear Research, Ukrainian Academy of Science, 03680 Kiev, Ukraine~$^{h}$} \\
$^{17}${\it J.~Stefan Institute, 1001 Ljubljana, Slovenia~$^{i}$} \\
$^{18}${\it University of Ljubljana, 1001 Ljubljana, Slovenia} \\
$^{19}${\it University of California, Los Angeles, CA 90024, USA~$^{j}$} \\
$^{20}${\it Lehrstuhl f\"ur Informatik V, Universit\"at Mannheim, D-68131 Mannheim, Germany} \\
$^{21}${\it University of Maribor, 2000 Maribor, Slovenia} \\
$^{22}${\it Institute of Theoretical and Experimental Physics, 117259 Moscow, Russia~$^{k}$} \\
$^{23}${\it Max-Planck-Institut f\"ur Physik, Werner-Heisenberg-Institut, D-80805 M\"unchen, Germany~$^{d}$} \\
$^{24}${\it Dept. of Physics, University of Oslo, N-0316 Oslo, Norway~$^{l}$} \\
$^{25}${\it Fachbereich Physik, Universit\"at Rostock, D-18051 Rostock, Germany~$^{d}$} \\
$^{26}${\it Fachbereich Physik, Universit\"at Siegen, D-57068 Siegen, Germany~$^{d}$} \\
$^{27}${\it Institute for Nuclear Research, INRNE-BAS, Sofia, Bulgaria} \\
$^{28}${\it Universiteit Utrecht/NIKHEF, 3584 CB Utrecht, The Netherlands~$^{a}$} \\
$^{29}${\it DESY, D-15738 Zeuthen, Germany} \\
$^{30}${\it Physik-Institut, Universit\"at Z\"urich, CH-8057 Z\"urich, Switzerland~$^{m}$} \\
$^{31}${\it visitor from Dipartimento di Energetica dell' Universit\`{a} di Firenze and INFN Sezione di Bologna, Italy} \\
$^{32}${\it visitor from P.N.~Lebedev Physical Institute, 117924 Moscow B-333, Russia} \\
$^{33}${\it visitor from Moscow Physical Engineering Institute, 115409 Moscow, Russia} \\
$^{34}${\it visitor from Moscow State University, 119899 Moscow, Russia} \\
$^{35}${\it visitor from Institute for High Energy Physics, Protvino, Russia} \\
$^{36}${\it visitor from High Energy Physics Institute, 380086 Tbilisi, Georgia} \\
$^\dagger${\it deceased}
\vspace{5mm}\\
$^{a}$ supported by the Foundation for Fundamental Research on Matter (FOM), 3502 GA Utrecht, The Netherlands \\
$^{b}$ supported by the CICYT contract AEN99-0483 \\
$^{c}$ supported by the German Research Foundation, Graduate College GRK 271/3 \\
$^{d}$ supported by the Bundesministerium f\"ur Bildung und Forschung, FRG, under contract numbers 05-7BU35I, 05-7DO55P, 05-HB1HRA, 05-HB1KHA, 05-HB1PEA, 05-HB1PSA, 05-HB1VHA, 05-HB9HRA, 05-7HD15I, 05-7MP25I, 05-7SI75I \\
$^{e}$ supported by the U.S. Department of Energy (DOE) \\
$^{f}$ supported by the Portuguese Funda\c c\~ao para a Ci\^encia e Tecnologia under the program POCTI \\
$^{g}$ supported by the Danish Natural Science Research Council \\
$^{h}$ supported by the National Academy of Science and the Ministry of Education and Science of Ukraine \\
$^{i}$ supported by the Ministry of Education, Science and Sport of the Republic of Slovenia under contracts number P1-135 and J1-6584-0106 \\
$^{j}$ supported by the U.S. National Science Foundation Grant PHY-9986703 \\
$^{k}$ supported by the Russian Ministry of Education and Science, grant SS-1722.2003.2, and the BMBF via the Max Planck Research Award \\
$^{l}$ supported by the Norwegian Research Council \\
$^{m}$ supported by the Swiss National Science Foundation
}

\abstract{
  The inclusive production cross sections of the strange
  vector mesons \kss, \kssb, and $\phi$ have been measured in
  interactions of \unit[920]{GeV} protons with C, Ti,
  and W targets with the \hb detector at the HERA storage ring.  
  Differential cross sections as a function of rapidity and transverse
  momentum have been measured in the central rapidity region and for
  transverse momenta up to $p_T=\unit[3.5]{GeV/c}$. The atomic number
  dependence is parametrised as $\sigma_{pA} = \sigma_{pN}*A^\alpha$,
  where $\sigma_{pN}$ is the proton-nucleon cross section. Within the
  phase space accessible, $\alpha(K^{*0}) =
  0.86 \pm 0.03
$,  
  $\alpha(\bar{K}^{*0}) =
  0.87 \pm 0.03
$, and $\alpha(\phi) = 
  0.96 \pm 0.02
$. 
  The total proton-nucleon cross sections, determined by extrapolating
  the differential measurements to full phase space, are  
  $\sigma_{pN\to  
    K^{*0}} = \unit[( 5.06 \pm 0.54
)]{mb}$,
  $\sigma_{pN\to \bar{K}^{*0}} =
  \unit[(4.02 \pm 0.45
)]{mb}$, and
  $\sigma_{pN\to \phi} =
  \unit[(1.17 \pm 0.11
)]{mb}$. For all resonances the Cronin
  effect is observed; 
  compared to the measurements of Cronin et al. for $K^\pm$ mesons,
  the measured values of $\alpha$ for $\phi$ mesons coincide with those 
  of $K^+$ mesons for all transverse momenta, while the enhancement
  for \kss/\kssb\ mesons is smaller. 
} 

\maketitle

\section{Introduction} The search for a deconfined state of quarks
and gluons in relativistic heavy ion collisions is a subject which has
received much attention in recent years \cite{won94,sat01}. It is a
challenging task to find a signal that enables the discrimination of the
short-lived locally deconfined state of quarks and gluons from a gas
of confined hadrons. An enhancement of the strange quark fraction has
been argued to appear in the quark gluon plasma \cite{let00}; hence one expects
an increase of strange hadron production in these processes. 

The rate of strange particle production depends on the temperature
at which the hadronisation step takes place, and on the chemical
potential. Hadronic interactions can influence the measurable strange
particle rate in a two-fold way \cite{mar02,raf01}: they can induce
transitions to other states, and, in the case of hadrons with a short
lifetime, rescattering of the final state particles can take place,
affecting the reconstructed invariant mass signal
\cite{mar05,jac05}. Since the strength 
of these final state interactions depends strongly on the lifetime,
\kss\ and $\phi$ mesons -- where the lifetimes differ by a
factor of about 10 -- have been proposed as a versatile tool to study
these effects in relativistic heavy ion reactions.

Consequently,
several new measurements of strange vector meson production have been
published recently by RHIC experiments. Both for \kss\ \cite{zha04,ada05}
and $\phi$ \cite{adl02,ada05b,adl05,adl05b,pal06,bly06,ma06}
mesons the influence of the very dense nuclear 
matter on the transverse momentum distributions and the line shape of
the resonances have been studied in detail. In contrast to this, only
limited information on $\phi$ production in proton-nucleus interactions
exists up to now. The $A$-dependence of $\phi$ production in
proton-nucleus interactions is studied in \cite{woe05}, while data in
dAu collisions are available from the STAR and PHENIX collaborations
\cite{bel05,cai05,pal06}. No data are available on \kss\ production.

Moreover, the study of the $p_T$ dependence of the inclusive production
rate of \kss\ and $\phi$ mesons in proton-nucleus (pA) interactions
is of interest in its own right. Cronin et al. \cite{cro75,cro77} observed a
nuclear enhancement of high-$p_T$ particles in pA as compared to pp
interactions for stable particles. In this paper we report the 
observation of the ``Cronin effect'' for the resonances \kss, \kssb,
and $\phi$. 
The effect has been ascribed by many models to a broadening of the
intrinsic transverse momentum of the primary partons in a hard
scattering process \cite{acc04}. The free parameters are fixed by
fitting the predictions to hard pp interactions. The high $p_T$ meson
spectra for proton-nucleus interactions are reasonably well described
by this ansatz, which however fails to reproduce the measured high
$p_T$ baryon spectra \cite{acc05}. On the other hand, the differences
between meson- and baryon-production in proton-nucleus reactions can
be explained by a final-state recombination model
\cite{hwa04}. Recently B.Z.~Kopeliovich et al. \cite{bzk02} have
formulated a parameter free 
description of the process. It explains the Cronin effect as a result
of a convolution of the initial parton distribution affected by
initial state interactions, followed by the hard 
scattering process and the fragmentation of the scattered
partons. From this ansatz, it follows that the steeper the $p_T$ dependence
of the cross section, the larger the Cronin effect should be, i.e. its
strength should depend on the flavor composition of the scattering
partons via its structure- and fragmentation functions dependence. 
Moreover, a weak energy dependence of the Cronin effect is expected
\cite{bzk02} in agreement with the observation of \cite{cro77} for
high-$p_T$ $K^{\pm}$ mesons produced in proton-nucleus interactions in
the energy range 200 GeV to 400 GeV. 

None of these models explicitly address the issue of the dependence of
the Cronin effect on the flavor or mass of the produced hadron.
The measurement presented here should motivate and aid further
theoretical efforts to understand the Cronin effect and thereby lead to
a deeper understanding of nuclear effects in pA collisions, particularly
in the area of strangeness production.

\begin{figure*}
  \begin{center}
    \subfigure[]{
      \includegraphics[clip,width=0.49\textwidth]
      {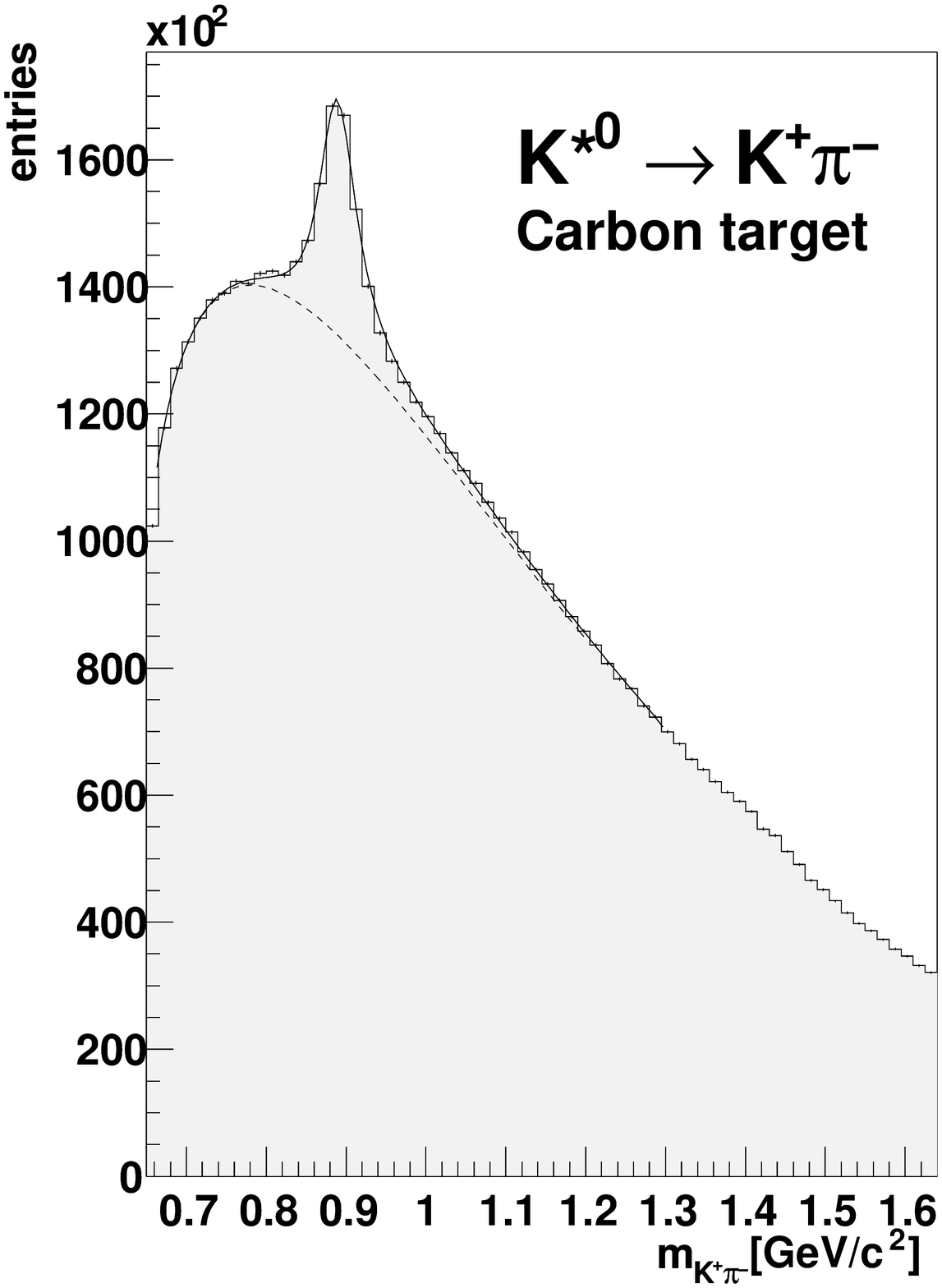}}     
    \subfigure[]{
      \includegraphics[clip,width=0.48\textwidth]
      {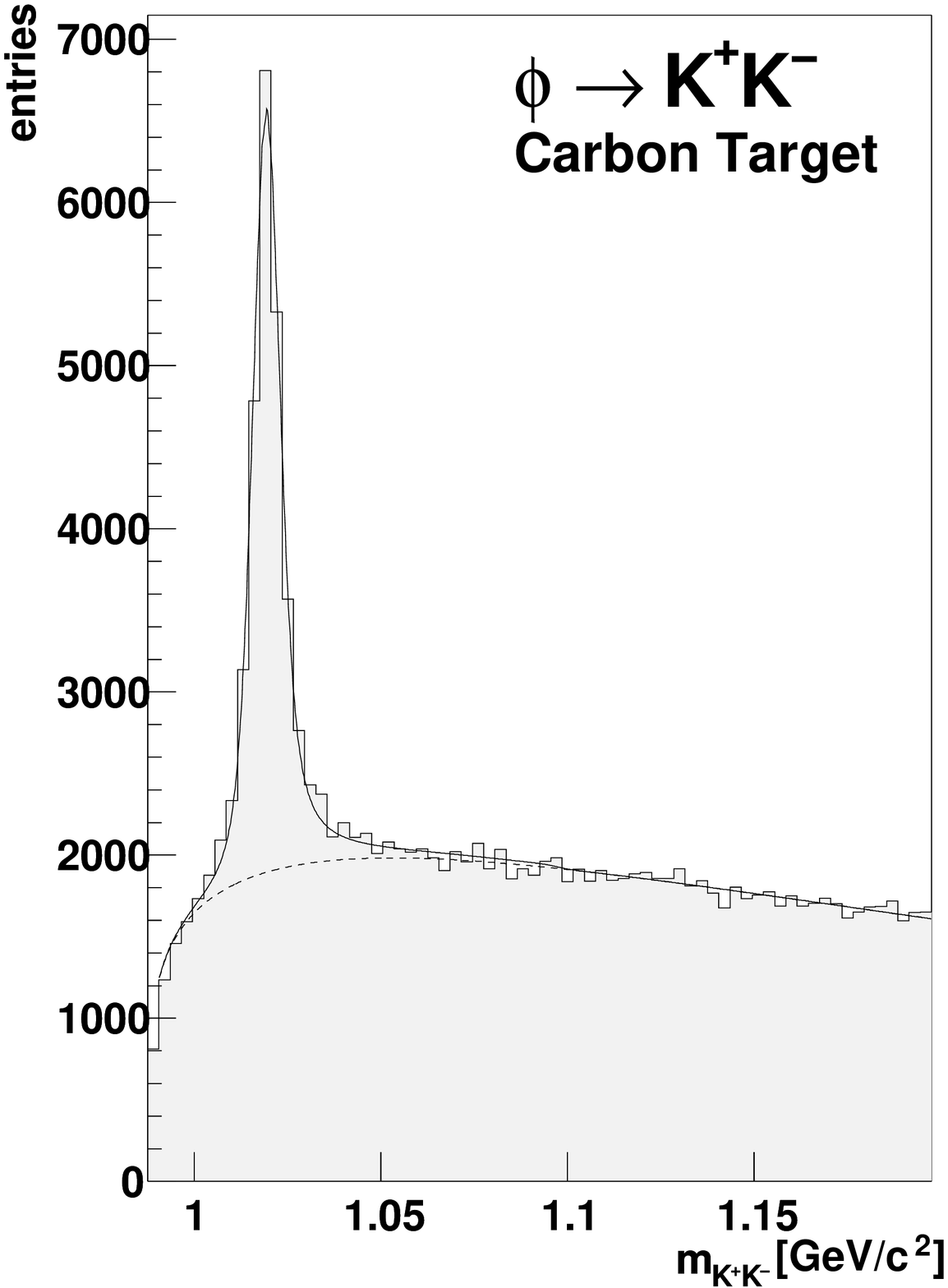}}
  \end{center}
\caption{Invariant mass distributions for a) $K^+\pi^-$ and b)
  $K^+K^-$ mass combinations on the carbon target after application of
  the cuts described in the text. The distributions were fit by a
  Gaussian-smeared relativistic Breit-Wigner function and a background
term (see text).} 
\label{fig:InvMass}       
\end{figure*}

\section{Data acquisition and event reconstruction}
The data were collected with the \hb detector operated at the proton
ring of the HERA storage ring complex. The pN center-of-mass energy was
$\sqrt{s}=\unit[41.6]{GeV}$. The \kss/\kssb\ data
cover the rapidity interval $-0.8\leq y\leq 0.3$ (in the
centre-of-mass system) and a transverse momentum range of $0\leq
p_T^2\leq \unit[12]{GeV^2/c^2}$. The $\phi$ results were obtained
for $-0.7\leq y\leq 0.25$ and $0.3\leq p_T^2\leq
\unit[12]{GeV^2/c^2}$. Here the lack of data for $p_T^2 <
\unit[0.3]{GeV^2/c^2}$ is due to the small opening angle of the $\phi$ 
decay in the lab system that makes the decay products escape through the
beam-pipe. 

The \hb\ detector is a forward magnetic
spectrometer with a large acceptance for particles produced in the
central region. The most important components of the detector for
this analysis are a wire target \cite{ehr04} inserted into the
halo of the \unit[920]{GeV} proton beam, a vertex detector
\cite{kno03} followed by a magnetic dipole field of \unit[2.13]{Tm},
a tracking system \cite{hoh01} using drift tube chambers, a Ring
Imaging Cherenkov detector (RICH)
\cite{ari04} which enables the identification of $K^\pm$ mesons 
above a threshold of \unit[9.6]{GeV/c}, and an electromagnetic
calorimeter \cite{zoc00}. A more detailed overview of the \hb detector
can be found elsewhere \cite{d0paper}.

The analysis is based on data recorded in the run period 2002/03. From a total of
$200\cdot 10^6$ events collected with three different target materials
a sub-sample of $132\cdot 10^6$ events taken in a 2-weeks period in December 2002
was selected to minimise systematic uncertainties. About $58\cdot
10^6$ events were recorded with a carbon (C) wire target, $21\cdot 10^6$ 
events with a titanium (Ti) target, and $53\cdot 10^6$ events with a tungsten 
(W) target. As the data were recorded at a moderate interaction rate of
$\unit[1.5\times 10^6]{s^{-1}}$, compared to an average rate of filled
proton bunches of $\unit[8.5\times 10^6]{s^{-1}}$ at the target, only
about 18\% of all bunch crossings with the target produce an inelastic
interaction. To reject empty bunch crossings an interaction trigger
accepted only those events, where either the  number of hits in the
RICH was $> 20$ (for a 
$\beta=1$ particle 32 photons are detected on average), or a cluster
with an energy $> \unit[1]{GeV}$ was detected in the inner part of the
electromagnetic calorimeter. The efficiency of the trigger was larger than
\unit[99]{\%}\cite{bru05}. In parallel to the interaction trigger,
events were selected in a random manner, regardless of whether or not
they contained an inelastic interaction. These random events
were used to study possible biases of the interaction trigger
\cite{bru05}. 

The events were reconstructed with the standard HE\-RA-B analysis package
\cite{abt02}, based on information from the vertex detector and the
tracking system. Charged 
particles were identified by matching the tracks with rings
reconstructed in the RICH. Using the track momentum information, the
likelihood probability for charged kaons, protons, light particles ($e$,
$\mu$, $\pi^\pm$), and a background hypothesis were calculated
\cite{ari04}.

The integrated luminosity was determined from the interaction rate at
the target monitored by various devices and from
the total inelastic cross section \cite{bru05}.

\section{Data analysis}
\subsection{$K^\pm$ meson identification and \kss\ ($\phi$)
  optimisation}
To identify \kss/\kssb\ and $\phi$, the decay channels \kssto, \kssbto,
and $\phi\to K^+K^-$ were investigated.
For all events with at least two oppositely charged tracks the
invariant mass $m_{\it inv}$ of the pair was separately calculated for
the $K^+\pi^-$, $K^-\pi^+$, and $K^+K^-$ mass assumption. For the
further analysis mass combinations were accepted if
\begin{equation*}
  \label{eq:kss_mass_constraint}
  \unit[0.63]{GeV/c^2} \leq m_{\it inv} \leq \unit[1.6]{GeV/c^2}
\end{equation*}
for the $K^\pm\pi^\mp$, and
\begin{equation*}
  \label{eq:phi_mass_constraint}
  \unit[0.988]{GeV/c^2} \leq m_{\it inv} \leq \unit[1.198]{GeV/c^2}
\end{equation*}
for the $K^+K^-$ combinations, respectively. In order to reduce the
combinatoric background, $K^\pm$ mesons were identified by the RICH.
For the \kss (\kssb) candidates, the kaon
identification was demanded for the positive (negative) track, while
for the $\phi$ candidates kaon identification was applied for both
tracks. Only $K^\pm$ candidates with a momentum $p_K \ge
\unit[9.7]{GeV/c}$ were considered, slightly above the kaon threshold
of the RICH. While for the determination of the $\phi$ signal a soft
likelihood probability cut of ${\cal L}_K \geq 0.3$ efficiently suppressed the
background, a harder cut of ${\cal L}_K \geq 0.95$ was required
for the \kss\ to reduce the combinatoric background. The non-existence
of mass reflections has been checked by experimental data and by the
Monte-Carlo (MC) simulation described in chapter \ref{sec:Acceptance}.

For momenta $p_K<\unit[40]{GeV/c}$ the efficiency for identification
of $K^\pm$ mesons is high, amounting to about \unit[90]{\%} \cite{ari04}. For
higher momenta the discrimination power drops and the
misidentification probability increases, 
since for such momenta the Che\-ren\-kov angles for $\pi$ and $K$
mesons become indistinguishable. The $K$ meson identification
efficiency as a function of $K$ momentum has been determined by MC.
It was checked with $\phi\to K^+K^-$ data \cite{sym04} by
identifying either one or both kaons and reconstructing the $\phi$
resonance. The difference in efficiency between data and MC of 5\% per
identified kaon was corrected for, and a residual systematic error of
2\% was estimated (see chapter \ref{sec:SysErrors}). A large fraction
of the produced \kss\ and $\phi$ candidates were rejected by the
momentum cut $p_K\geq  \unit[9.7]{GeV/c}$. The total efficiency for
$K$ mesons from \kss\ and $\phi$ resonance decays to survive the RICH
identification and the momentum cut was about \unit[15]{\%}.  

The significance of the signal was improved further by applying
soft track quality cuts. In case of the vertex detector at least 6
hits per track were demanded.
In the tracking system a cut of at least 15 hits per track was used to
enhance the significance of the \kss\ ($\phi$) signal without reducing
the efficiency. On average, a track comprised 12 hits in the vertex
detector and 40 hits in the tracking system, respectively.  
The \kss\ signal significance was improved further by
applying a cut of $p_\pi\geq \unit[2]{GeV/c}$ to pion track candidates
\cite{eld04}. The cuts were optimised for significance using signal
MC and real data background events. 

\subsection{Signal determination}
Typical invariant mass distributions for the $K^+\pi^-$ and $K^+K^-$
combinations are shown in Fig. \ref{fig:InvMass}, 
after application of the cuts mentioned above. To facilitate the
extraction of the 
signal, the resonance was described by a relativistic $P$-wave
Breit-Wigner \cite{jac64}:
\begin{eqnarray}
  \label{eq:relBW}
  BW(m) &=& \frac{m\cdot m_0\cdot\Gamma(m)}{(m_0^2 - m^2)^2 +
    (m_0\cdot\Gamma(m))^2}\\
  \Gamma(m) &=& \Gamma_0\cdot \left(\frac{q}{q_0}\right)^{2l+1}
  \cdot \frac{m_0}{m}, 
\end{eqnarray}
with
\begin{eqnarray*}
  \label{eq:relBW2}
  \Gamma_0 &=& \Gamma(m_0)~\text{the natural width}\\
  q(m)_{|K^{*0}} &=&
  \frac{\sqrt{(m^2-(m_K+m_\pi)^2)(m^2-(m_K-m_\pi)^2)}}{2m}\\
  q(m)_{|\phi} &=&
  \frac{\sqrt{m^2-4m_K^2}}{2}\\
  q_0 &=& q(m_0)\\
  l &=& 1.
\end{eqnarray*}
$q(m)$ is the momentum of the decay products in the rest frame of the
mother particle, and $l$ is the angular momentum transferred.
To take into account effects of the detector resolution the
Breit-Wigner distribution was numerically folded with a Gaussian of
width $\sigma_m$ in the fits of the measured mass spectra. It has been
checked that the mass resolutions in experimental
data and MC are consistent within errors. The mass $m_0$
of the resonance and the width $\sigma_m$ were free parameters of the
fit, while the natural width $\Gamma_0$ of the resonance was fixed to
its PDG value \cite{pdg04}. The number of resonance decays was
estimated by integrating the fitted signal in the interval $\pm
3\Gamma_0$ around the resonance mass $m_0$. A variation of the
fit interval by $\Gamma_0$ changed the yield up to 5\%. The same was
observed for the MC used for acceptance correction, thus compensating 
the effect seen in data.

The background was parametrised by expressions which have been applied
successfully in other experiments. For the \kss/\kssb\ the
parametrisation of \cite{agu91} was adopted: 
\begin{equation}
  \label{eq:BgParKstar}
  BG_{K^{*0}}(m) = p_0 \cdot \left(\frac{q}{m}\right)^{p_1} \cdot 
  e^{-p_2q~-~p_3 q^2},
\end{equation}
where $p_0\dots p_3$ are free parameters and $q$ is the momentum
transfer introduced above. For the $\phi$ meson the background was
described by a parametrisation proposed in \cite{gra78}:
\begin{equation}
  \label{eq:BGParPhi}
  BG_\phi(m) = p_0 \cdot m_{\it rel}^{p_1} \cdot e^{-p_2 m~-~p_3m^2},
\end{equation}
where $m_{\it rel} = m - 2m_K$ and $p_0\dots p_3$ again are free
parameters of the fit. 
As shown in Fig. \ref{fig:InvMass}, these combinations of signal and
background functions describe the data properly.

\begin{table*}
  \caption{Number of inelastic events ($N_{\it ia}$), number of
    \kss, \kssb, and 
    $\phi$ candidates ($N_{\it VM}$) after analysis 
    cuts, reconstructed mass ($m_{\it rec}$), and experimental mass
    resolution ($\sigma_{\it m}$) for each target material. Errors are
  statistical only.}  
  \label{tab:StatisticsSummary}
  \centering
  \begin{tabular}{lcccc}
    \hline\noalign{\smallskip}
    & $N_{\it ia}$ & $N_{\it VM}$ & $m_{\it rec}$ [MeV/$c^2$] &
    $\sigma_{\it m}$ [MeV/$c^2$]\\
    \hline\noalign{\smallskip}
    & \multicolumn{4}{c}{$pA\to K^{*0}X$}\\
    \hline\noalign{\smallskip}
    C & $58\cdot 10^6$ & $187574\pm 1726$ & $890.5\pm 0.2$ & $5.1\pm
    0.2$\\
    Ti & $21\cdot 10^6$ & $83231\pm 1237$ & $890.7\pm 0.4$ & $5.0\pm
    0.3$\\
    W & $53\cdot 10^6$ & $262356\pm 2420$ & $890.8\pm 0.2$ & $5.3\pm
    0.3$\\
    \hline\noalign{\smallskip}
    & \multicolumn{4}{c}{$pA\to \bar{K}^{*0}X$}\\
    \hline\noalign{\smallskip}
    C & $58\cdot 10^6$ & $149721\pm 1622$ & $890.8\pm 0.2$ & $4.6\pm
    0.1$\\  
    Ti & $21\cdot 10^6$ & $65823\pm 1147$ & $890.7\pm 0.4$ & $5.1\pm
    0.4$\\
    W & $53\cdot 10^6$ & $195373\pm 2225$ & $891.5\pm 0.3$ & $4.8\pm
    0.2$\\
    \hline\noalign{\smallskip}
    & \multicolumn{4}{c}{$pA\to \phi X$}\\
    \hline\noalign{\smallskip}
    C & $58\cdot 10^6$ & $15379\pm 204$ & $1019.2\pm 0.1$ & $2.6\pm 0.1$\\
    Ti & $21\cdot 10^6$ & $7580\pm 142$ & $1019.2\pm 0.1$ & $2.7\pm 0.1$\\
    W & $53\cdot 10^6$ & $26971\pm 295$ & $1019.3\pm 0.1$ & $2.6\pm 0.1$\\
    \hline\noalign{\smallskip}
  \end{tabular}
\end{table*}

\begin{table*}
  \caption{Summary of systematic errors for the integrated and total
    production cross section measurements. See text for a more
    detailed explanation. Numbers are grouped target material
    wise. The last two errors are due 
    to the phase space extrapolation and have been applied only to the total
    cross sections. Numbers are given in \%.}
  \label{tab:systematics}
  \begin{center}
    \begin{tabular}{l c c c | c c c | c c c} 
      \hline\noalign{\smallskip}
      & \multicolumn{3}{c}{C} & \multicolumn{3}{c}{Ti} &
      \multicolumn{3}{c}{W}\\
      \hline\noalign{\smallskip}
      & \kss & \kssb & $\phi$ & \kss & \kssb & $\phi$ & \kss & \kssb & $\phi$\\
      \hline\noalign{\smallskip}
      MC statistics & 1.0 & 1.3 & 1.6 & 1.0 & 1.3 & 1.1 & 1.0 & 1.3 &
      0.7 \\ 
      Bg. parametrisation & 2.2 & 1.9 & - & 1.7 & 2.3 & - & 2.5 & 2.3 & - \\
      Cut variation & 2.8 & 2.8 & 6.2 & 2.8 & 2.8 & 7.0 & 2.8 & 2.8 & 7.1 \\
      BR & 0.1 & 0.1 & 1.22 & 0.1 & 0.1 & 1.22 & 0.1 & 0.1  & 1.22 \\
      Luminosity & & $4.8$ & & & $5.0$ & & & $3.8$ \\
      Luminosity scaling & & $2.0$ & & & $2.0$ & & & $2.0$ \\
      Track efficiency & & 3.0 & & & 3.0 & & & 3.0 \\
      \hline\noalign{\smallskip}
      $p_T^2$-extr. & - & - & 6.6 & - & - & 8.8 & - & - & 8.2 \\
      $y$-extr. & 2.6 & 4.1 & 6.3 & 2.6 & 4.5 & 6.0 & 2.6 & 4.5 & 4.9\\
      \hline\noalign{\smallskip}
    \end{tabular}
  \end{center}
\end{table*}

It has been shown \cite{sym04,eld04} that both the fitted masses and the
detector resolution $\sigma_m$ derived from the data were stable for
the whole data taking period and independent of the target material.
Also the resonance yields, normalized to the
integrated luminosity, were stable during the whole data taking period.
Table \ref{tab:StatisticsSummary} summarises the \kss, \kssb\ and
$\phi$ statistics obtained after cuts, the reconstructed mass
positions, and the experimental mass resolutions.

The bin size for the $p_T^2$- and rapidity ($y$-) distributions
were chosen to be broader than the resolution
determined from MC simulation. From MC we find
that in each bin more than 80\% of the reconstructed vector
mesons were generated in the same bin.

\subsection{Acceptance correction\label{sec:Acceptance}}
Acceptance correction functions were determined from MC events
using FRITIOF 7.02 \cite{pi92}. The detector simulation was
based on GEANT 3.21 \cite{gea93} and includes the measured
hit resolutions, mapping of inefficient channels, and electronic
noise. The simulated events were processed by the same reconstruction
chain as the data. Since the FRITIOF
generator uses an approximation for the generation of hard
parton-parton scattering, the MC $p_T^2$ distributions show a steeper
slope than observed in the data. To make sure that the reconstruction
efficiencies and the influence of the analysis cuts are correctly
described by the simulation, the MC events were reweighted
in $p_T^2$ according to 
the data for both the \kss/\kssb\ and the $\phi$ meson. 
It was checked that after reweighting both the rapidity and azimuthal
MC distributions match those of the data.

The total efficiency for each $p_T^2$- and $y$-bin was determined with
the help of the reweighted MC events by comparing the number of
reconstructed and produced \kss/$\phi$ mesons in each bin. The number
of reconstructed mesons was obtained similar to data by a fit to the
reconstructed invariant mass spectra. Typical
efficiencies vary between 5\% and 30\% \cite{sym04,eld04}. Studies of
systematic uncertainties due to signal extraction and background
estimation show that they are negligible compared to the
statistical error in the corresponding interval.
Note that for each rapidity bin the efficiencies are
determined by integrating over the measured $p_T^2$-range and vice
versa. 

\subsection{Systematic errors \label{sec:SysErrors}}

Possible sources of systematic uncertainties have been studied in
detail:
Since especially the \kss/\kssb\ invariant mass distributions are
affected by a large combinatorial background peaking close to the
resonance signal, an alternative background parametrisation to
(\ref{eq:BgParKstar}) was used, which was proven to be successful in
studies of \kss\ production in $e^+e^-$-collisions \cite{lin92}: 
\begin{equation}
  \label{eq:BgParaSys}
  BG(m) = p_0~\left[1-\left(\frac{1}{m+p_1}\right)^{p_2}\right] \cdot
  \left(1+p_3 m\right),
\end{equation}
where $p_0\dots p_3$ are free parameters. The signal yields
obtained in most $p_T^2$- and $y$-intervals agree for the two
analyses within their statistical uncertainty. The absolute difference
of the yields, derived using the two background parametrisations, is
considered as a systematic error. It amounts to about 2\% for the
integrated cross sections. For the $\phi$ meson a corresponding
study \cite{sym04} provided much smaller deviations, and their
contribution to the systematic error can be safely neglected.

The variation of the analysis cuts, including the uncertainties in the
kaon identification efficiency (2.0\% per identified particle), resulted
in a systematic uncertainty of 2.8\% for the \kss/\kssb\ and 6.2-7.1\%
for the $\phi$ analysis (depending on the target material). The
systematic error of the luminosity consists of an overall scaling
error of 2\% and a target material dependent part of 3.8-5.0\%, and
the systematic 
error of the branching ratios \cite{pdg04} amounts to 0.1\% and 1.22\%
for \ksstop/\kssbto and $\phi\to K^+K^-$, respectively. A systematic
error on the track efficiency of 1.5\% per track was estimated using
$K^0_s\to \pi^+\pi^-$ decays \cite{per04}. A
summary of all systematic errors is given in Table 
\ref{tab:systematics}.

\section{Experimental results}
\subsection{Differential cross sections \label{sec:DiffCS}}
Due to limited statistics, we did not calculate the
double-differential cross section, but rather
determined the one-dimensional rates $dN/dp_T^2$ and $dN/dy$, which were
integrated over the measured $y$ and $p_T^2$ range, respectively.
Given the corresponding one-dimensional efficiencies\\ $\epsilon_{\it
  acc}(p_T^2)$ and $\epsilon_{\it acc}(y)$, the differential cross
sections can be written as:
\begin{equation}
  \label{eq:DiffCS}
  \frac{d\sigma}{dx} = \frac{1}{BR\cdot{\cal L}}
  \frac{dN}{dx} \frac{1}{\epsilon_{\it acc}(x)},
\end{equation}
where $x = \left\{y, p_T^2\right\}$. $BR$ and ${\cal L}$ denote the
branching 
ratio of the decay and the integrated luminosity of the data set,
respectively. The acceptance boundaries of the measurement in
rapidity and transverse momentum squared are
given in Table \ref{tab:cross_sections}. 
\begin{figure*}
  \begin{center}
    \subfigure[]{
      \includegraphics[clip,width=0.49\textwidth,bb=0 0 530 650]
      {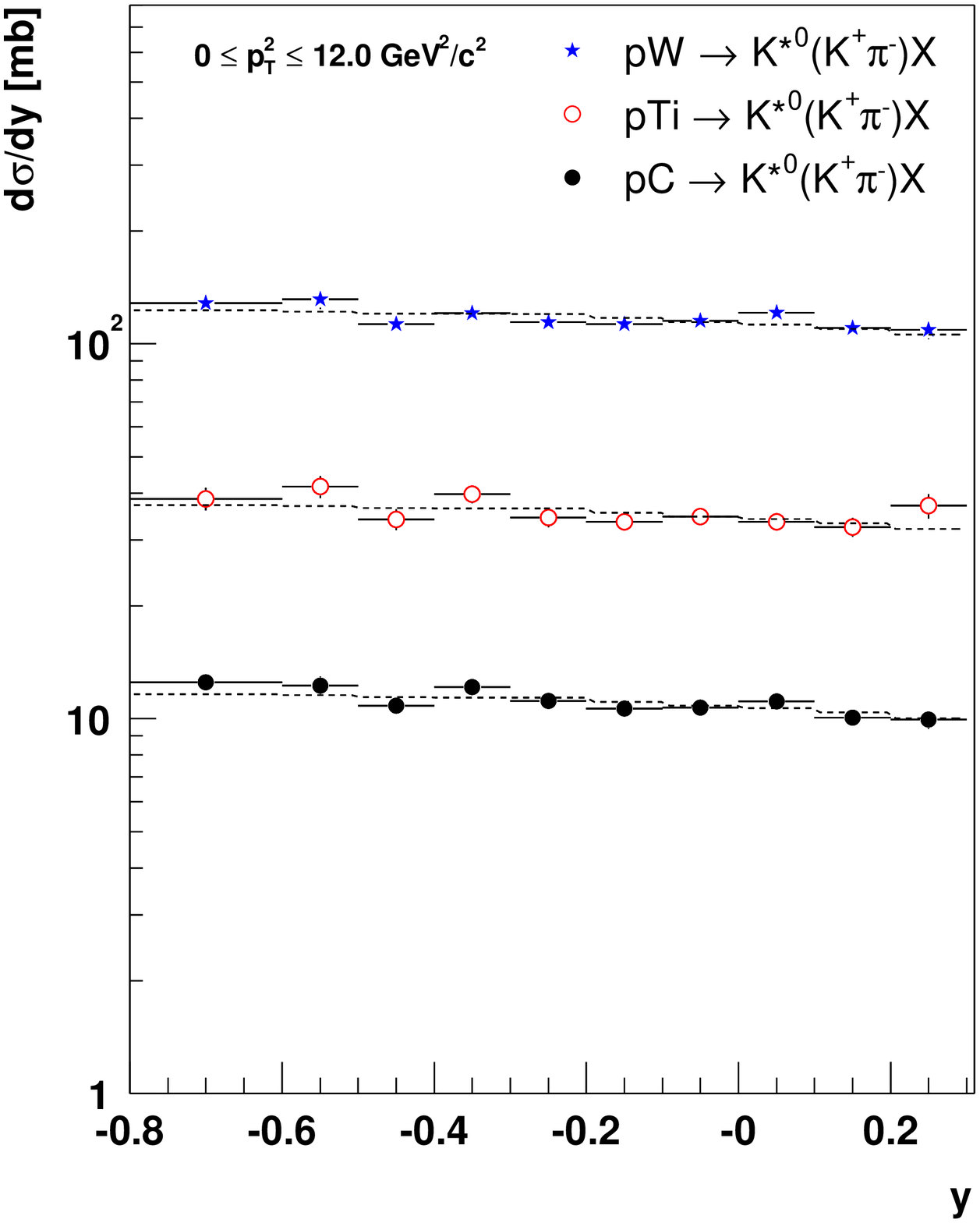}}
    \subfigure[]{
      \includegraphics[clip,width=0.49\textwidth,bb=0 0 530 650]
      {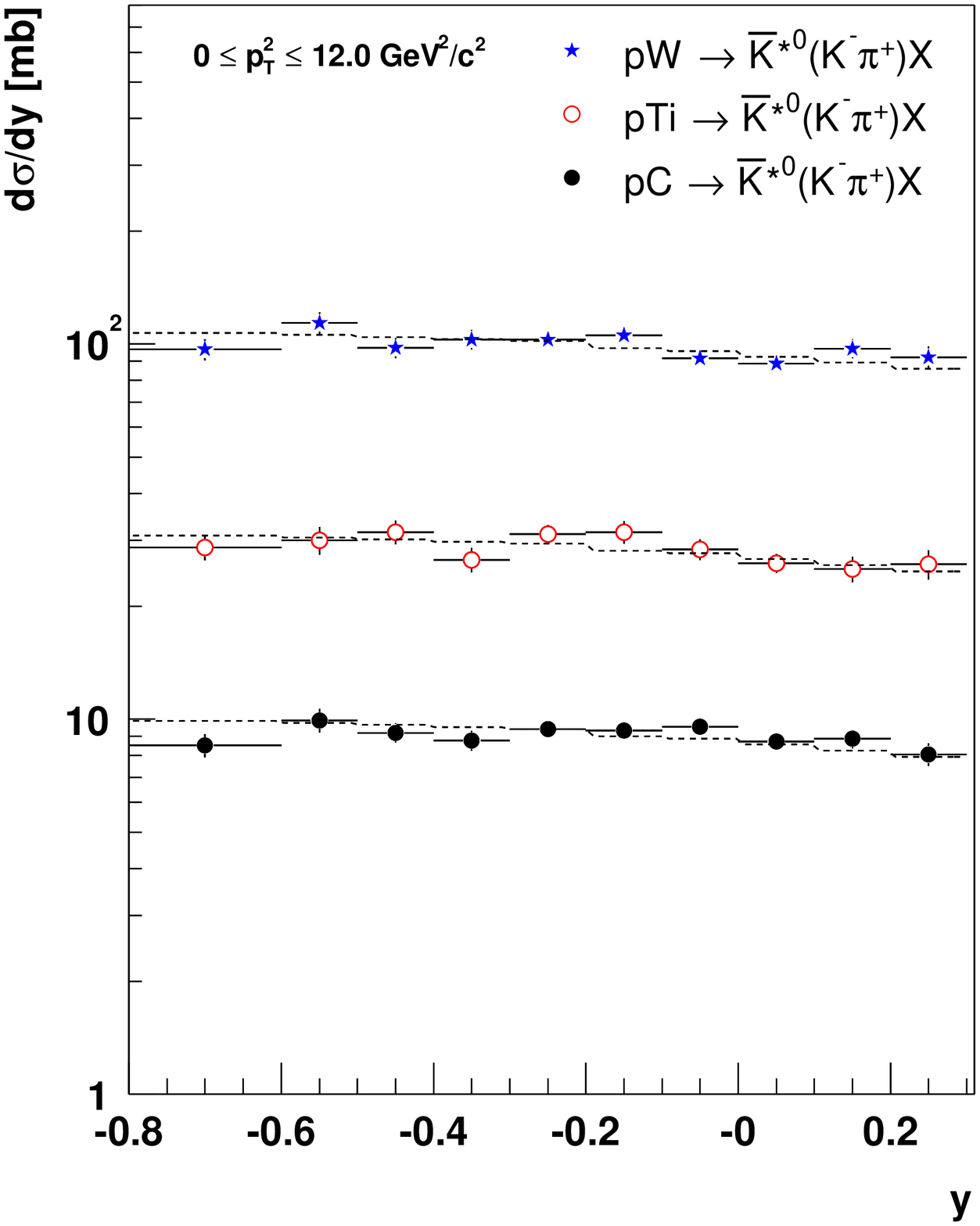}}
  \end{center}
  \caption{Measured inclusive differential cross sections
    $d\sigma/dy$ as a function of rapidity for a) \kss\
    and b) \kssb\ meson 
    production in the range $0\leq p_T^2\leq \unit[12]{GeV^2/c^2}$. 
    Horizontal bars indicate the widths of the individual bins, while
    vertical errors bars represent the uncertainties due to 
    statistics. Note that for most bins the error is smaller than the
    size of the symbols. The dashed lines show the
    (scaled) results of the FRITIOF MC.}
  \label{fig:kstar_rap}
\end{figure*}

\begin{figure}
  \includegraphics[clip,width=0.49\textwidth, bb=0 0 530 650]
  {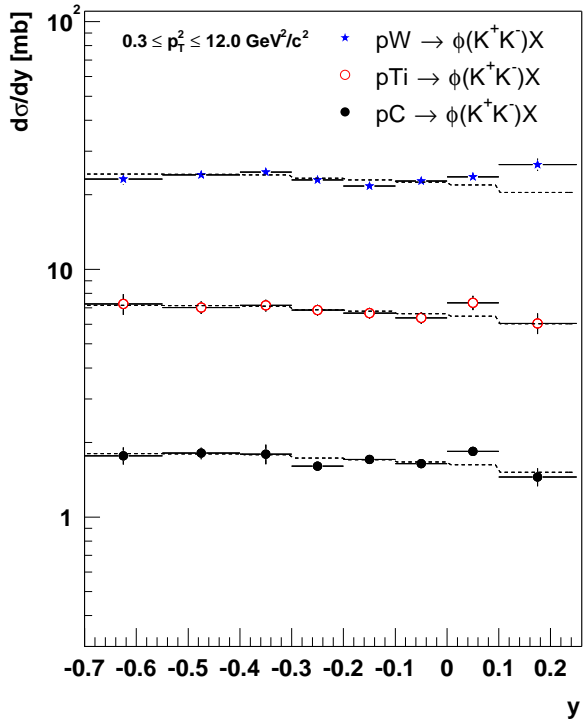}
  \caption{Measured inclusive differential cross section
    $d\sigma/dy$ as a function of rapidity for $\phi$
    meson 
    production in the range $\unit[0.3]{GeV^2/c^2}\leq p_T^2\leq
    \unit[12]{GeV^2/c^2}$. The vertical error bars reflect the
    uncertainties due to statistics.
    The dashed lines show the (scaled) results of the FRITIOF
    MC.} 
  \label{fig:phi_rap}
\end{figure}

The measured cross sections for the \kss\ and the \kssb\ mesons
as a function of rapidity are shown in Fig. \ref{fig:kstar_rap}. The
cross sections are rather flat as expected in the central 
rapidity region \cite{fey69}. The \kssb\ cross sections are
suppressed in comparison to the \kss\ cross sections by about 20\%,
in agreement with the results of the FRITIOF MC. The rapidity
distributions for $\phi$ mesons produced on C, Ti, and W targets are
shown in Fig. \ref{fig:phi_rap}. Again the cross section in the
central region is flat.  

\begin{figure*}
  \begin{center}
    \subfigure[]{
      \includegraphics[clip,width=0.49\textwidth, bb=0 0 530 650]
      {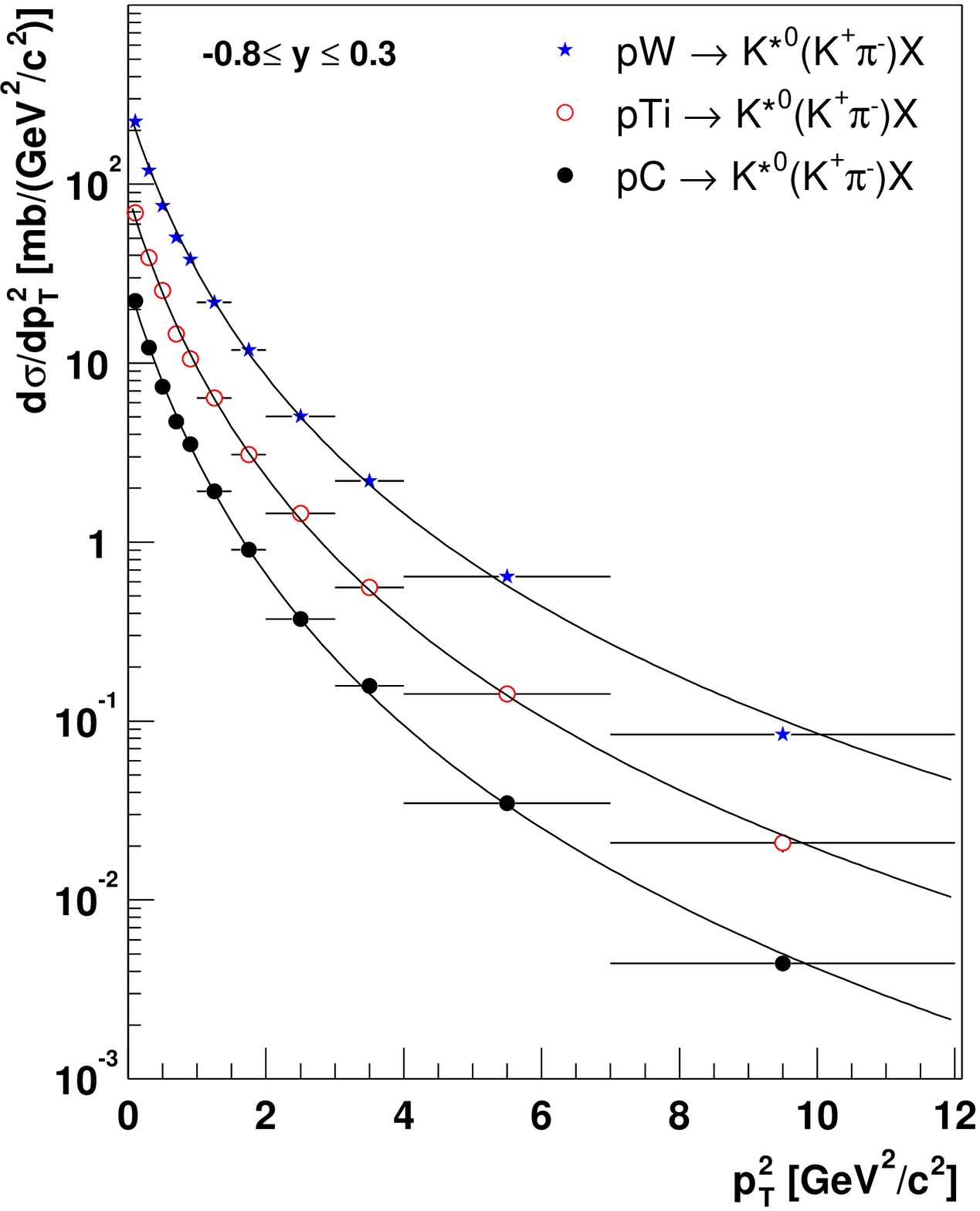}}
    \subfigure[]{
      \includegraphics[clip,width=0.49\textwidth, bb=0 0 530 650]
      {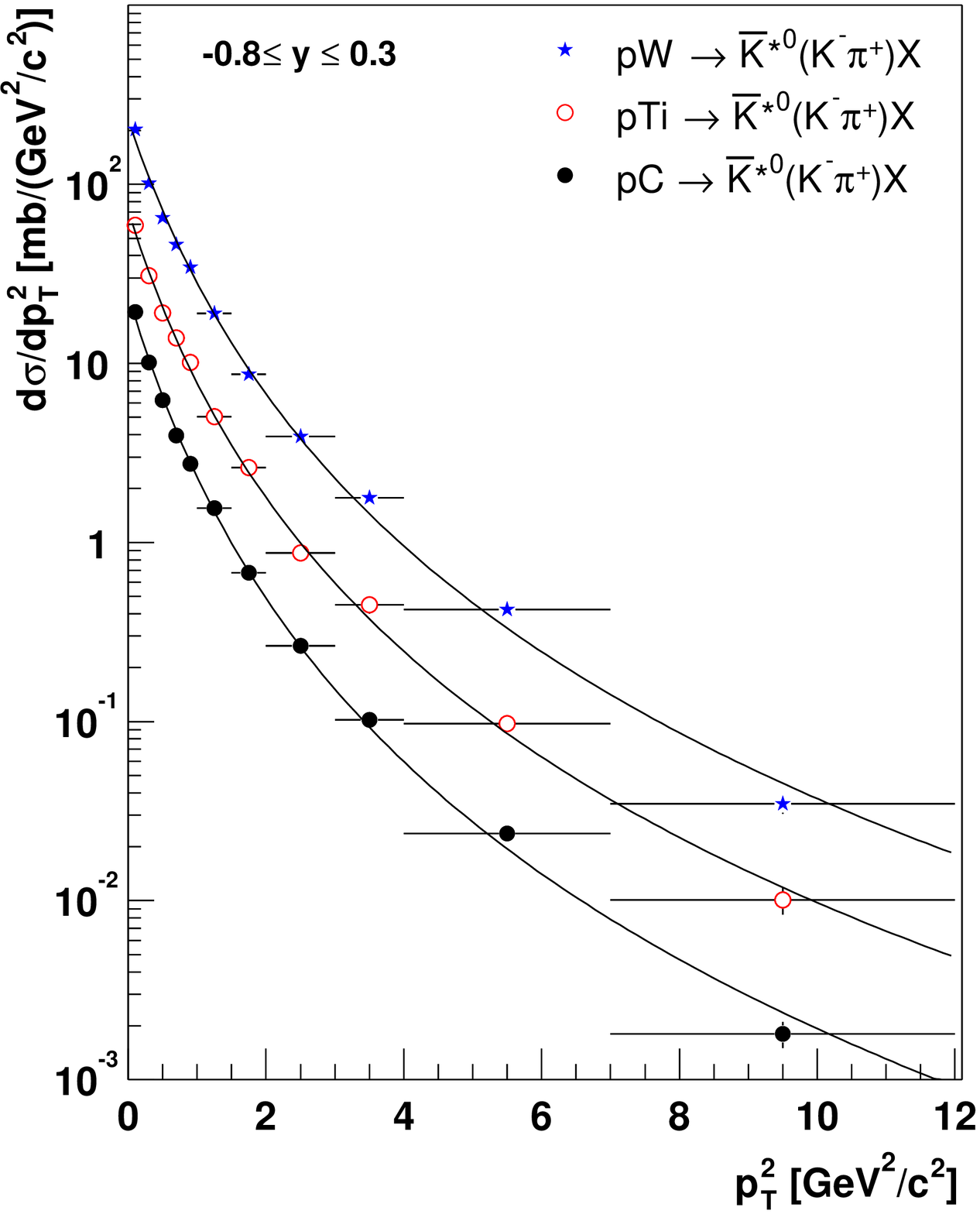}}
  \end{center}
  \caption{Measured inclusive differential cross sections
    $d\sigma/dp_T^2$ as a function of $p_T^2$ for a) \kss\
    and b) \kssb\ 
    production. The vertical error bars reflect statistical errors
    only. Fits to the parametrisation (\ref{eq:PowerLaw_pt2}) are
    superimposed.} 
  \label{fig:kstar_pt2}
\end{figure*}

Transverse momentum spectra are shown in
Fig. \ref{fig:kstar_pt2} for the \kss\ and \kssb.
The spectra deviate from a simple exponential shape
at large transverse momenta, which can be traced back to hard
parton-parton scattering \cite{gei90}. Common exponential
parametrisations like
\begin{equation}
  \label{eq:Expo_pt2}
  \frac{d\sigma}{dp_T^2} = C_2~e^{-ap_T^2},
\end{equation}
or
\begin{equation}
  \label{eq:Expo_mt}
  \frac{d\sigma}{dp_T^2} = C_3~e^{-bm_T},~m_T=\sqrt{m_0^2+p_T^2}
\end{equation}
only fit the data for $p_T^2\lesssim \unit[1]{GeV^2/c^2}$.

On the contrary, as is shown in Fig. \ref{fig:kstar_pt2}, a power-law
parametrisation,  
\begin{equation}
  \label{eq:PowerLaw_pt2}
  \frac{d\sigma}{dp_T^2} = C_1\left(1 + \frac{p_T^2}{p_0^2}\right)^{-\beta},
\end{equation}
describes the spectra over the full $p_T^2$ range of the
measurements. The $p_T^2$-spectrum of \kssb\ mesons, which have no
valence quarks in common with nucleons, is steeper than the \kss\ meson  
spectrum. Similar observations have been made in hard pp collions for
$K^-$ mesons in comparison to $K^+$ mesons \cite{bre84,gei90}. In the
latter case this observation has been traced back to the fact that 
$K^-$ mesons are, for the kinematic condition of the  experiment,
dominantly gluon fragments \cite{bre85,gei90}. QCD calculations
\cite{abr85} show that for a fixed value of $x_T=2~p_T/\sqrt{s}$ the
relative contribution of gluon fragments to $K^-$ (\kssb) production
decreases while the sea quark share is enhanced when the CMS energy
is increased.  

\begin{figure}
  \includegraphics[clip,width=0.49\textwidth, bb=0 0 530 650]
  {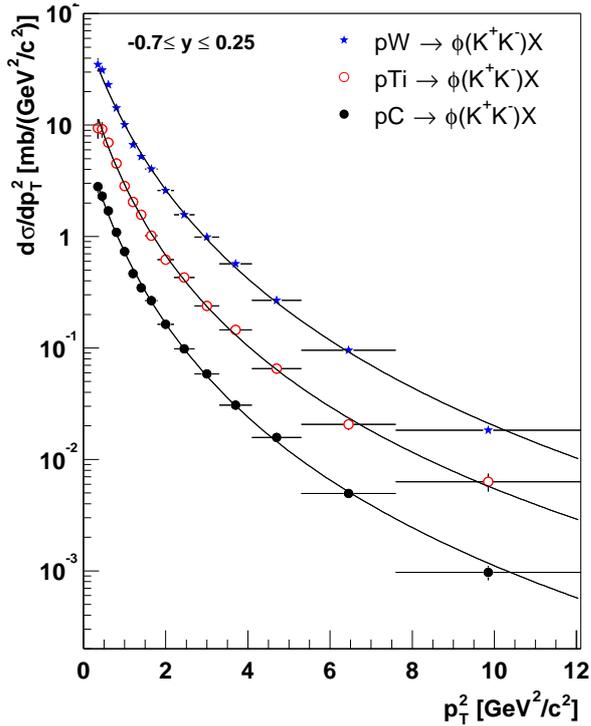}
  \caption{Measured inclusive differential cross sections
    $d\sigma/dp_T^2$ as a function of $p_T^2$ for $\phi$
    meson 
    production. The vertical error bars represent the uncertainties due to
    statistics. Fits to the parametrisation (\ref{eq:PowerLaw_pt2})
    are superimposed.} 
\label{fig:phi_pt2}
\end{figure}

The same parametrisation (\ref{eq:PowerLaw_pt2}) successfully fits the 
transverse momentum distributions for the $\phi$, shown in
Fig. \ref{fig:phi_pt2}. The suppression of $\phi$ mesons as compared to
\kssb mesons by a factor of 3 to 4 is compatible with the known
strange\-ness suppression in fragmentation processes \cite{lin92}. No
simple qualitative explanation exists for the observation that the
\kssb $p_T$ spectrum is steeper than that of $\phi$ meson.

The fit parameters for all particles are collected in Table
\ref{tab:fit_params}. Note that the $\chi^2$ of the
tungsten fits is considerably worse than those of the other target
materials. This is likely caused by an enhancement of background due
to the large multiplicity of the pW interactions. The $\chi^2$
improves however considerably when leaving out the bin of largest $p_T^2$
in the fits. The differential cross sections $d\sigma/dy$
and $d\sigma/dp_T^2$ for \kss, \kssb, and $\phi$ production are
summarised in Tables \ref{tab:CrossSectionSummary} and
\ref{tab:CrossSectionSummary2}. 

\begin{table*}
  \begin{center}
  \caption{Results of the power-law fits (\ref{eq:PowerLaw_pt2}) for
    the differential cross sections $d\sigma/dp_T^2$. Fits were
    performed within the full $p_T^2$-range of the measurements and
    in the rapidity interval accessible by the experiment ($-0.8\leq
    y\leq 0.3$ for \kss/\kssb, $-0.7\leq y\leq 0.25$ for 
    $\phi$). The last column gives the mean $p_T$ of the cross
    sections, calculated from the fit parameters.}
  \label{tab:fit_params}
  \begin{tabular}{l c c c c c }
    \hline\noalign{\smallskip}
    & $\unit[C_1]{[mb/GeV^2]}$ & $\beta$ & $\unit[p_0^2]{[GeV^{2}/c^2]}$ 
    & $\chi^2/\text{n.d.f.}$ & $\unit[\langle p_T \rangle]{[GeV/c]}$\\
    \hline\noalign{\smallskip}
    & \multicolumn{5}{c}{$pA\to K^{*0}X$}\\
    \hline\noalign{\smallskip}
    C & $ 28.5 \pm 1.1 $ & $ 4.17 \pm 0.12 $ & $ 1.36 \pm 0.08 $ & 9.1/8 & $0.66 \pm 0.07$ \\

    Ti & $ 87.1 \pm 5.2 $ & $ 3.89 \pm 0.14 $ & $ 1.30 \pm 0.11 $ & 8.3/8 & $0.69 \pm 0.09$ \\

    W & $ 268.5 \pm 10.9 $ & $ 3.75 \pm 0.10 $ & $ 1.32 \pm 0.08 $ & 24.5/8 & $0.72 \pm 0.07$ \\

    \hline\noalign{\smallskip}
    & \multicolumn{5}{c}{$pA\to \bar{K}^{*0}X$}\\
    \hline\noalign{\smallskip}
    C & $ 24.4 \pm 1.0 $ & $ 4.67 \pm 0.16 $ & $ 1.52 \pm 0.11 $ & 14.2/8 & $0.64 \pm 0.07$ \\

    Ti & $ 71.9 \pm 4.4 $ & $ 4.39 \pm 0.21 $ & $ 1.51 \pm 0.14 $ & 13.0/8 & $0.67 \pm 0.11$ \\

    W & $ 230.5 \pm 10.2 $ & $ 4.51 \pm 0.16 $ & $ 1.68 \pm 0.12 $ & 36.9/8 & $0.69 \pm 0.08$ \\

    \hline\noalign{\smallskip}
    & \multicolumn{5}{c}{$pA\to \phi X$}\\
    \hline\noalign{\smallskip}
    C & $ 7.6 \pm 0.8 $ & $ 4.04 \pm 0.15 $ & $ 1.26 \pm 0.13 $ & 12.6/12 & $0.66 \pm 0.09$ \\

    Ti & $ 31.9 \pm 5.0 $ & $ 3.81 \pm 0.21 $ & $ 1.15 \pm 0.17 $ & 10.5/12 & $0.66 \pm 0.14$ \\

    W & $ 73.9 \pm 6.1 $ & $ 4.20 \pm 0.14 $ & $ 1.65 \pm 0.14 $ & 28.4/12 & $0.72 \pm 0.09$ \\

    \hline\noalign{\smallskip}
  \end{tabular}
  \end{center}
\end{table*}

\subsection{Integrated inclusive and total cross sections}
\begin{table*}
  \begin{center}
  \caption{Extrapolated total cross sections $\sigma_{\it pA}$, and
    integrated inclusive differential production cross sections
    $\sigma_{\it vis}$ 
    within the accessible phase space of the measurement. The errors
    denote the statistical error 
    and the combined systematic error, respectively. The rapidity and $p_T^2$
    coverage of the measurements are given in columns 3 and
    4. For the $\phi$ meson, $R_{p_T^2}^{\it extr}$ denotes
    the factor used for extrapolation of the integrated cross sections
    to $p_T^2 = 0$. $\Delta\sigma_{\it vis}$ is the fraction of the
    total cross section within
    the rapidity coverage of the measurement, based on the FRITIOF
    \cite{pi92} and HIJING \cite{wan91} Monte-Carlo generators. Note
    that for these numbers the $p_T^2$-extrapolation for the $\phi$
    cross sections was applied beforehand.}
  \label{tab:cross_sections}
  \begin{tabular}{l c c c c c c}
    \hline\noalign{\smallskip}
    & $\sigma_{\it pA}$ [mb]
    & $\sigma_{\it vis}$ [mb]
    & rapidity interval & $p_T^2$ range [GeV$^2/c^2$] 
    & $R_{p_T^2}^{\it extr}$
    & $\Delta\sigma_{\it vis} [\%]$\\
    \hline\noalign{\smallskip}
    & \multicolumn{5}{c}{$pA\to K^{*0}X$}\\
    \hline\noalign{\smallskip}
    C & 
     $ 43.9 \pm 0.6\pm 3.3 $ 
&
     $ 12.1 \pm 0.2 \pm 0.9 $ 
 &
    -0.8 -- +0.3 & 0.0 -- 12.0 &
    1 & $27.4\pm 0.7$\\
    Ti & 
     $ 141.2 \pm 2.6\pm 10.6 $ 
&
     $ 38.5 \pm 0.7 \pm 2.7 $ 
 & & &
    1 & $27.3\pm 0.7$\\
    W &  
     $ 465.9 \pm 6.4\pm 32.7 $ 
&
     $ 127.5 \pm 1.7 \pm 8.3 $ 
 & & &
    1 & $27.4\pm 0.7$\\
    \hline\noalign{\smallskip}
    & \multicolumn{5}{c}{$pA\to \bar{K}^{*0}X$}\\
    \hline\noalign{\smallskip}
    C & 
     $ 36.0 \pm 0.6\pm 2.9 $ 
&
     $ 10.0 \pm 0.2 \pm 0.7 $ 
&
    -0.8 -- +0.3 & 0.0 -- 12.0 &
    1 & $27.7\pm 1.1$\\ 
    Ti &
     $ 111.5 \pm 2.5\pm 9.7 $ 
&
     $ 31.1 \pm 0.7 \pm 2.3 $ 
 & & &
    1 & $27.9\pm 1.3$ \\
    W & 
     $ 388.8 \pm 6.9\pm 30.8 $ 
&
     $ 107.6 \pm 1.9 \pm 7.0 $ 
 & & &
    1 & $27.7\pm 1.3$\\
    \hline\noalign{\smallskip}
    & \multicolumn{5}{c}{$pA\to \phi X$}\\
    \hline\noalign{\smallskip}
    C & 
     $ 11.0 \pm 0.2\pm 1.4 $ 
&
     $ 1.60 \pm 0.03 \pm 0.14 $ 
&
    -0.7 -- +0.25 & 0.3 -- 12.0&
    $1.94\pm 0.13$ & $28.3\pm 1.8$ \\
    Ti &
     $ 44.7 \pm 1.0\pm 6.0 $ 
&
     $ 6.4 \pm 0.1 \pm 0.6 $ 
 & & & 
    $1.97\pm 0.17$ & $28.3\pm 1.7$ \\
    W & 
     $ 135.1 \pm 1.6\pm 17.0 $ 
&
     $ 21.9 \pm 0.3 \pm 2.0 $ 
& & &
    $1.72\pm 0.14$ & $27.9\pm 1.4$ \\
    \hline\noalign{\smallskip}
  \end{tabular}
 \end{center}
\end{table*}

The integrated cross sections in the acceptance region of the
\hb detector are calculated by fitting
the shape of the MC distributions to the differential cross sections
$d\sigma/dy$. The results of the fits are shown as dashed lines
in Figs. \ref{fig:kstar_rap} and \ref{fig:phi_rap}. 
The integrated cross sections  $\sigma_{\it vis}$ in
the phase space accessible to the \hb  experiment are listed in Table
\ref{tab:cross_sections}. Note that integrating the $p_T^2$
differential distributions yields consistent cross sections. 

The good agreement of the \kss\ and \kssb\ transverse momentum spectra
with the power-law parametrisation (\ref{eq:PowerLaw_pt2}) for small
$p_T$ suggests that this is true also for the $\phi$ differential
cross sections, where only data with $p_T^2\geq \unit[0.3]{GeV^2/c^2}$
are available. Hence we have used this para\-me\-tri\-sation to extrapolate
the $\phi$ differential cross section to $0\leq p_T^2\leq
\unit[0.3]{GeV^2/c^2}$ when calculating the total cross section.
The extrapolation factors are summarised in Table
\ref{tab:cross_sections}. The systematic uncertainty of the
extrapolations is estimated by using parametrisation
(\ref{eq:Expo_pt2}) as an alternative and taking the difference
between the two approaches. Leaving out the bin of largest $p_T^2$ in
the power-law fit increases the extrapolated $\phi$ cross section by
about 6\%, which is fully compatible with this systematic
uncertainty. 

Taking into account the limited rapidity coverage of the detector, the
integrated cross sections are then extrapolated to full phase space
using target-specific correction factors $\Delta\sigma_{\it
  vis}$. These factors are derived from the MC simulators FRITIOF
\cite{pi92} and HIJING \cite{wan91}, and the average of both was taken
for extrapolation. The systematic error was estimated by taking half
the difference between the FRITIOF and the HIJING results. As
quantified in Table \ref{tab:cross_sections} about 28\% of the
total cross section is accessible to the  experiment. Correcting
the limited phase space coverage leads to the total
inclusive cross sections 
\begin{equation}
\sigma_{pA} = \frac{R^{\it extr}_{p_T^2}\cdot\sigma_{\it vis}}
{\Delta\sigma_{\it vis}}
\end{equation}
for \kss, \kssb, and $\phi$ production
in pC-, pTi-, and pW-interactions (collected in Table
\ref{tab:cross_sections}). The systematic error of the 
cross section extrapolation in total amounts to about 2.6\% for the
\kss, 4.5\% for the \kssb, and 10.0\% for the $\phi$ total
production cross sections (Table \ref{tab:systematics}).

\begin{table}
  \caption{Atomic mass number dependence of the integrated and total
    production cross sections, and extrapolated production cross
    sections in proton-nucleon reactions.
    The errors of $\alpha$ reflect the statistical errors and
    all target-material dependent systematic errors.
    For $\sigma_{pN}$, full errors due to statistics and 
    systematics are given. Errors were added in quadrature.}
  \label{tab:alpha}
  \centering
  \begin{tabular}{l c c c}
    \hline\noalign{\smallskip}
    & $pN\to K^{*0}X$ & $pN\to \bar{K}^{*0}X$ & $pN\to \phi X$ \\
    \hline\noalign{\smallskip}
    $ \alpha_{\it vis} $ 
    & $0.86 \pm 0.03
$ 
    & $0.87 \pm 0.03
$ 
    & $0.96 \pm 0.02
$ \\
    $ \alpha_{pA}$  
    & $0.87 \pm 0.03
$ 
    & $0.87 \pm 0.03
$
    & $0.91 \pm 0.02
$ \\
    \hline\noalign{\smallskip}
    $\unit[\sigma_{pN}]{[mb]} $  
    & $ 5.06 \pm 0.54
$ 
    & $4.02 \pm 0.45
$ 
    & $1.17 \pm 0.11
$\\
    \hline\noalign{\smallskip}
  \end{tabular}
\end{table}

\subsection{Atomic mass number dependence of the integrated and total
  cross sections} 

\begin{figure}
  \begin{center}
    \subfigure[]{
      \includegraphics[clip,width=0.4\textwidth]
      {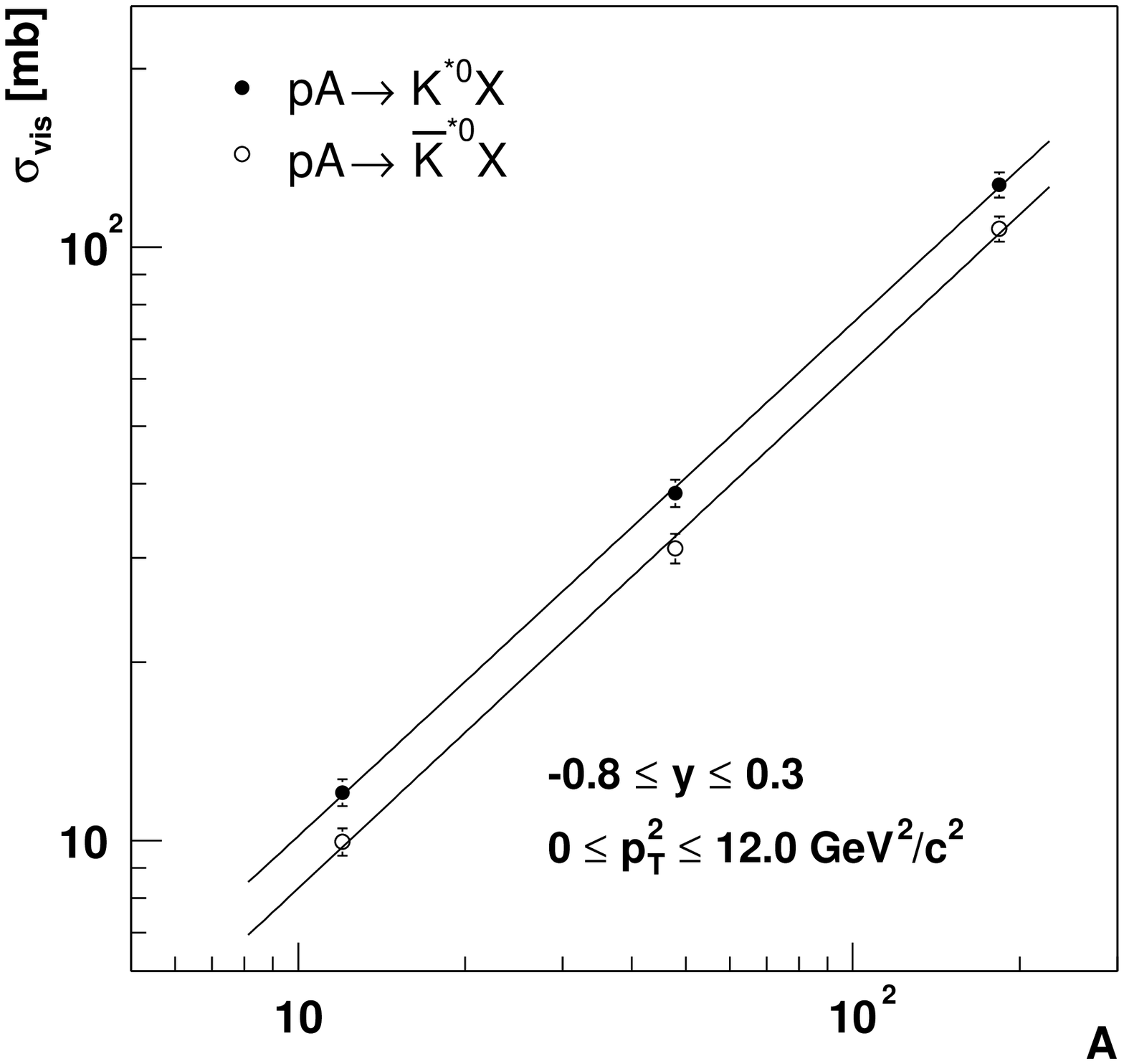}}
    \subfigure[]{
      \includegraphics[clip,width=0.4\textwidth]
      {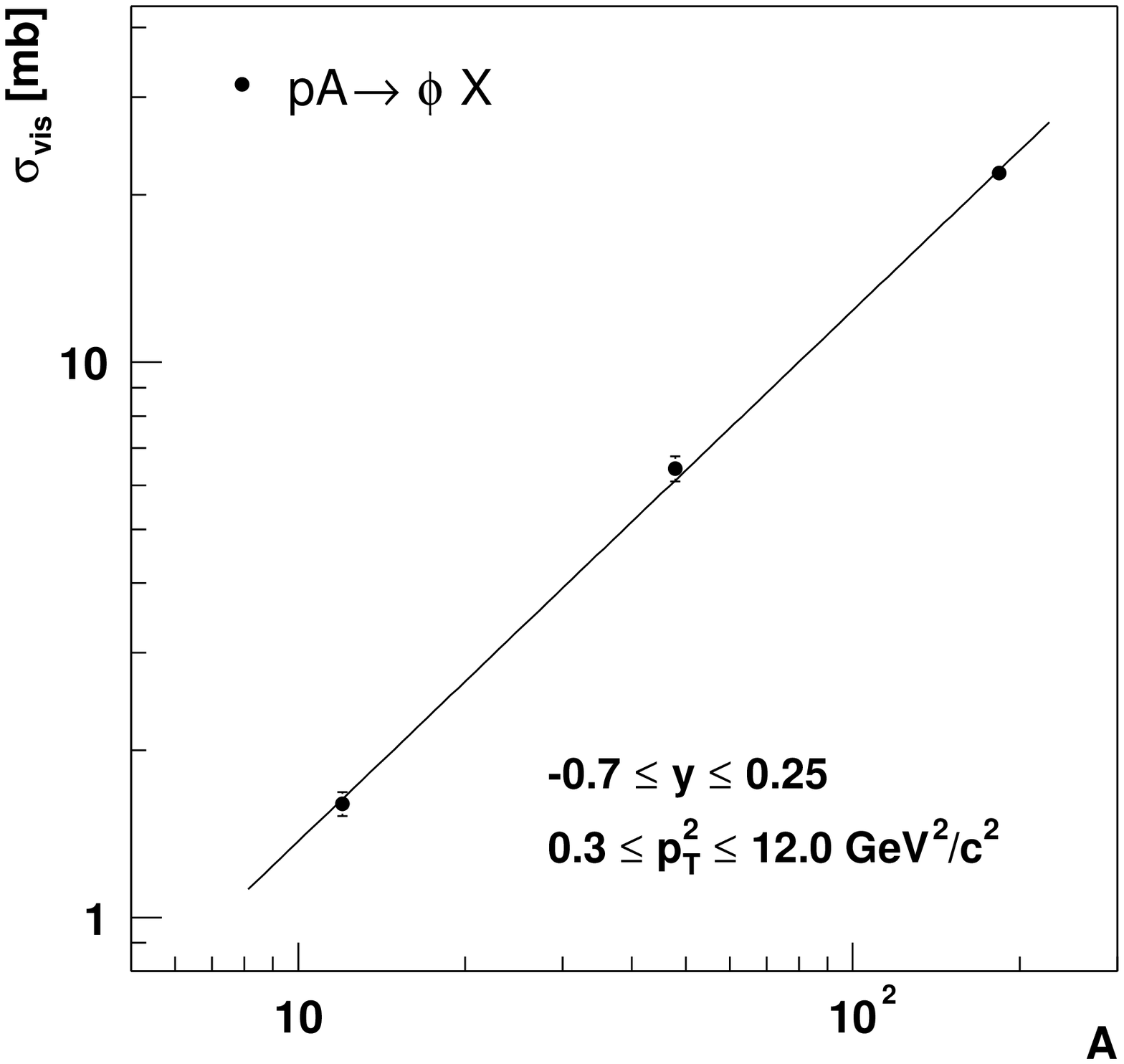}}
  \end{center}
  \caption{Atomic mass number dependence of the integrated inclusive cross
  sections for a) \kss/\kssb\ and b) $\phi$ production. Fits to the
  parametrisation (\ref{eq:alpha}) are superimposed. The errors shown
  reflect the statistical error as well as all target-material dependent
  systematic errors, added in quadrature.} 
\label{fig:A_dep}
\end{figure}

Fig. \ref{fig:A_dep} shows the atomic mass number dependence of the
integrated cross sections $\sigma_{\it vis}$ for the production of
\kss, \kssb, and $\phi$ mesons in pA-interactions. The atomic
number dependence is well-described by a power-law
\begin{equation}
  \label{eq:alpha}
  \sigma_{\it vis} = \sigma_{\it vis,0}\cdot A^{\alpha_{\it vis}},
\end{equation}
where $\sigma_{\it vis,0}$ is the visible proton-nucleon cross section.
The results for the exponent $\alpha_{\it vis}$ are collected in Table
\ref{tab:alpha}. Comparing the values with those for inelastic
pA cross sections, $\alpha_{\it inel} = 0.71\pm 0.01$ \cite{gei91}, it
follows that strange vector meson production shows a stronger
$A$-dependence than the total inelastic cross section. While the latter
scales roughly with the cross-sectional area of the nucleus, the \kss\
and $\phi$ production cross sections show a tendency towards scaling
with the volume. 

\begin{figure}
  \begin{center}
    \subfigure[]{
      \includegraphics[clip,width=0.49\textwidth, bb=0 0 520 360]
      {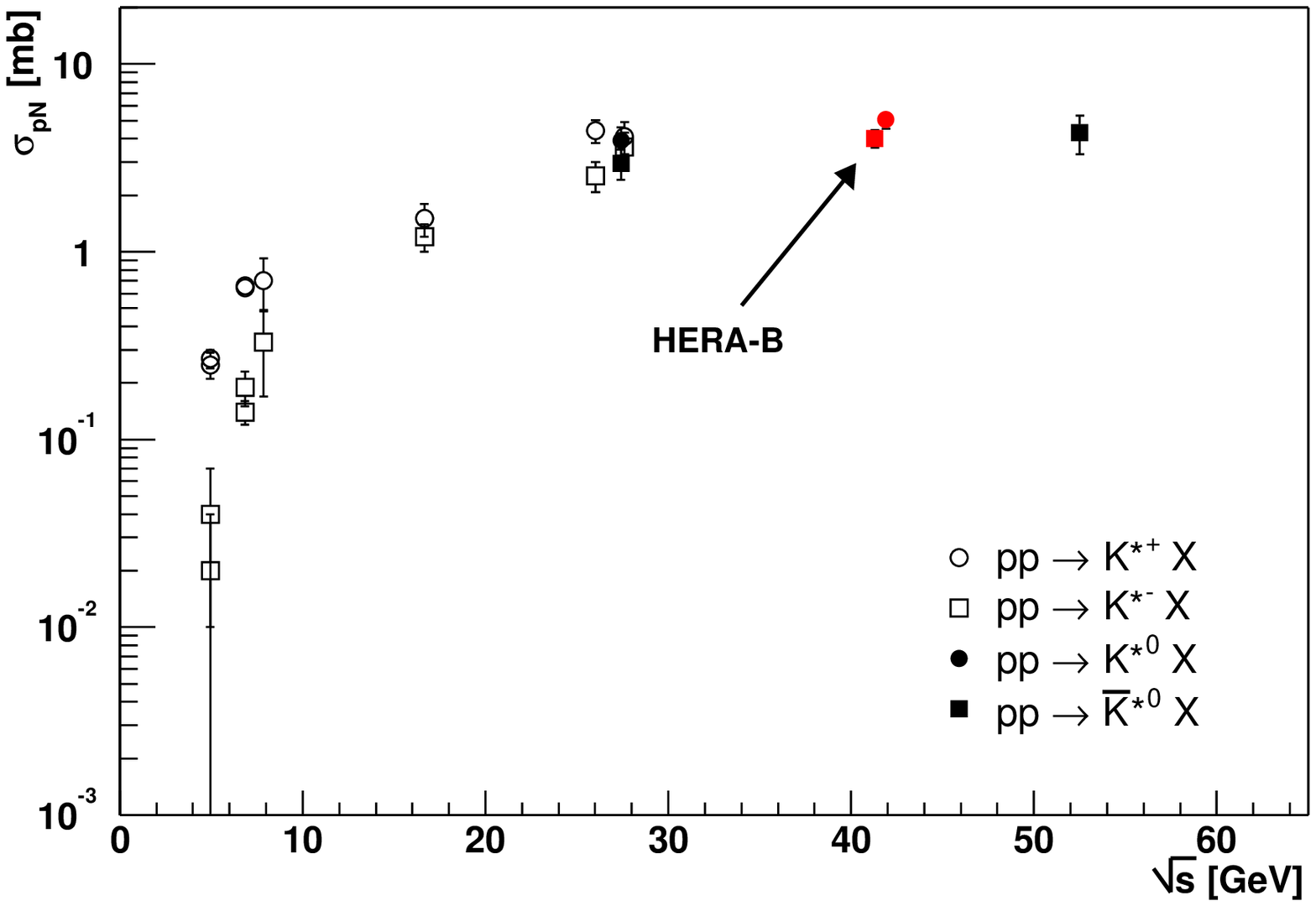}}
    \subfigure[]{
      \includegraphics[clip,width=0.49\textwidth, bb=0 0 520 360]
      {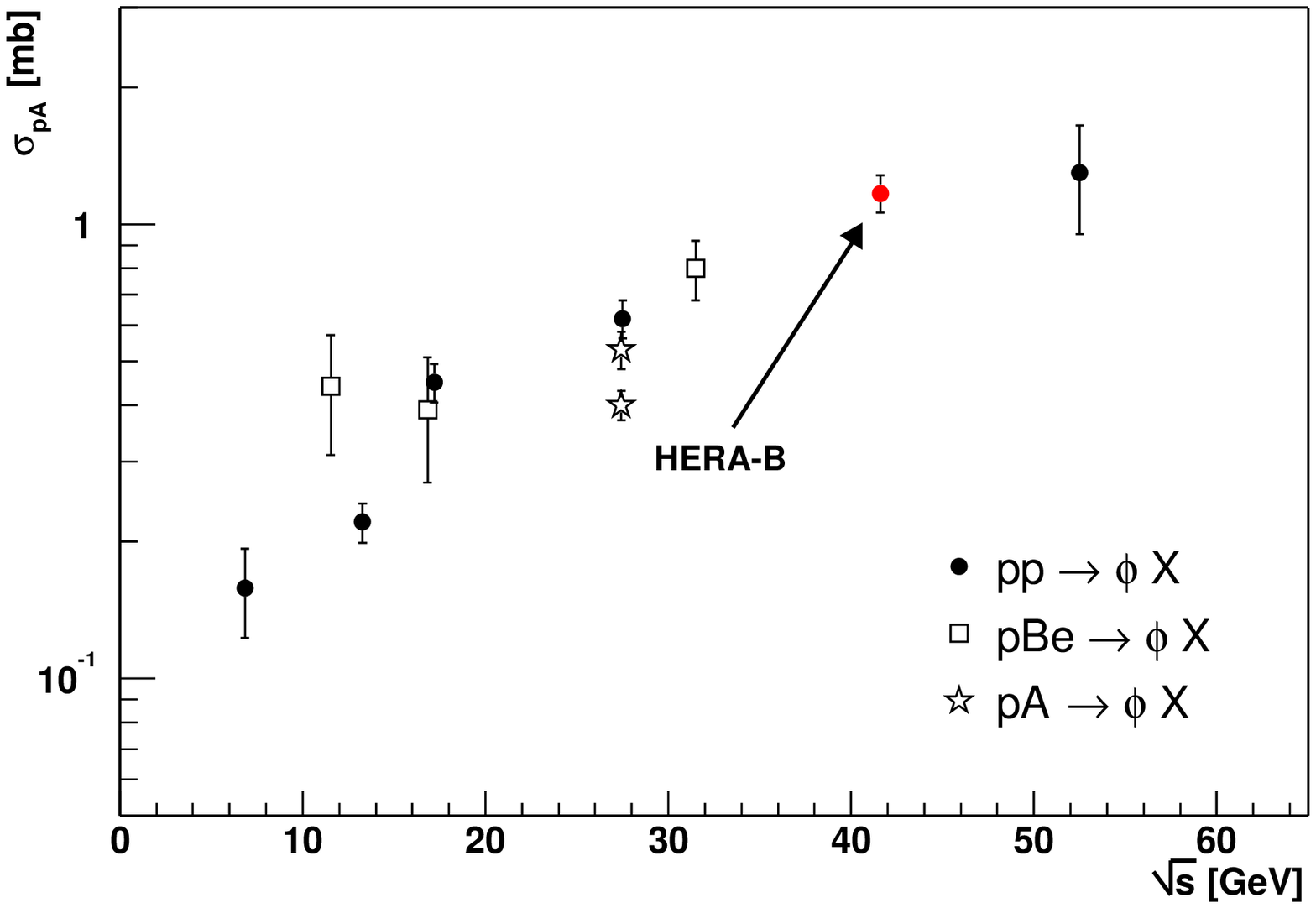}}
  \end{center}
  \caption{a) Comparison of \kss/\kssb\ and $K^{*\pm}$ meson total
    production 
    cross sections at various centre-of-mass energies
    \cite{blo74,boc79,kic79,dri81,bri82,azi86,bog88,agu91}.  The
    results of 
    this analysis have been slightly displaced from each other for
    better visibility. b) same as a) but for $\phi$ meson production
    \cite{agu91,sam01,kar91,dau81,dri81,blo75,ant82,and76,woe05}. 
    Results of the pBe measurements have been extrapolated to $A=1$
    using $\alpha=0.91\pm 0.02$ from \cite{woe05}.}  
  \label{fig:pN_comp}
\end{figure}

The same parametrisation can be used to describe the $A$-dependence 
of the cross sections extrapolated to the full phase space,
\begin{equation}
  \label{eq:alpha2}
  \sigma_{pA} = \sigma_{pN}\cdot A^{\alpha_{pA}}.
\end{equation}
$\sigma_{pN}$ is an estimate of the total cross section for strange
vector meson production in proton-nucleon reactions. The fitted values
for $\alpha_{pA}$ and $\sigma_{pN}$ are included in Table \ref{tab:alpha},
and the results for $\sigma_{pN}$ are compared to those measured at other
energies in pp and pA reactions (Fig. \ref{fig:pN_comp}). 
Recently, the NA60 collaboration measured the production of $\phi$ mesons
in collisions of 400 GeV protons with various nuclei
\cite{woe05}. Fitting their data to eq. \ref{eq:alpha2}, they obtain
$\alpha=0.91\pm0.02$, which is in good agreement with our
measurements. Consequently, the pBe data displayed in
Fig.~\ref{fig:pN_comp}~b) were rescaled using this result.  

\subsection{Cronin effect}
\begin{figure}
  \begin{center}
    \subfigure[]{
      \includegraphics[clip,width=0.4\textwidth]
      {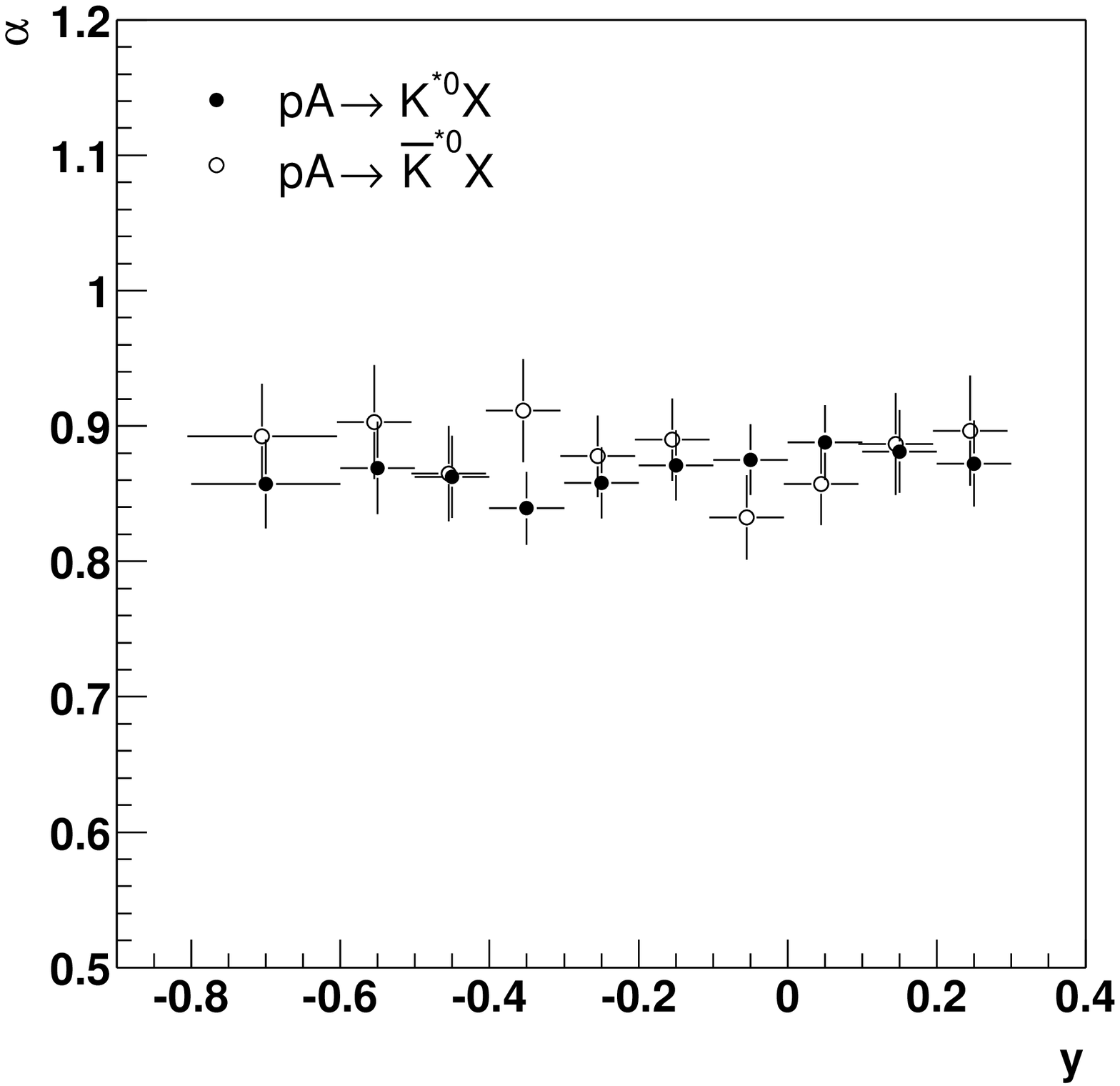}}
    \subfigure[]{
      \includegraphics[clip,width=0.4\textwidth]
      {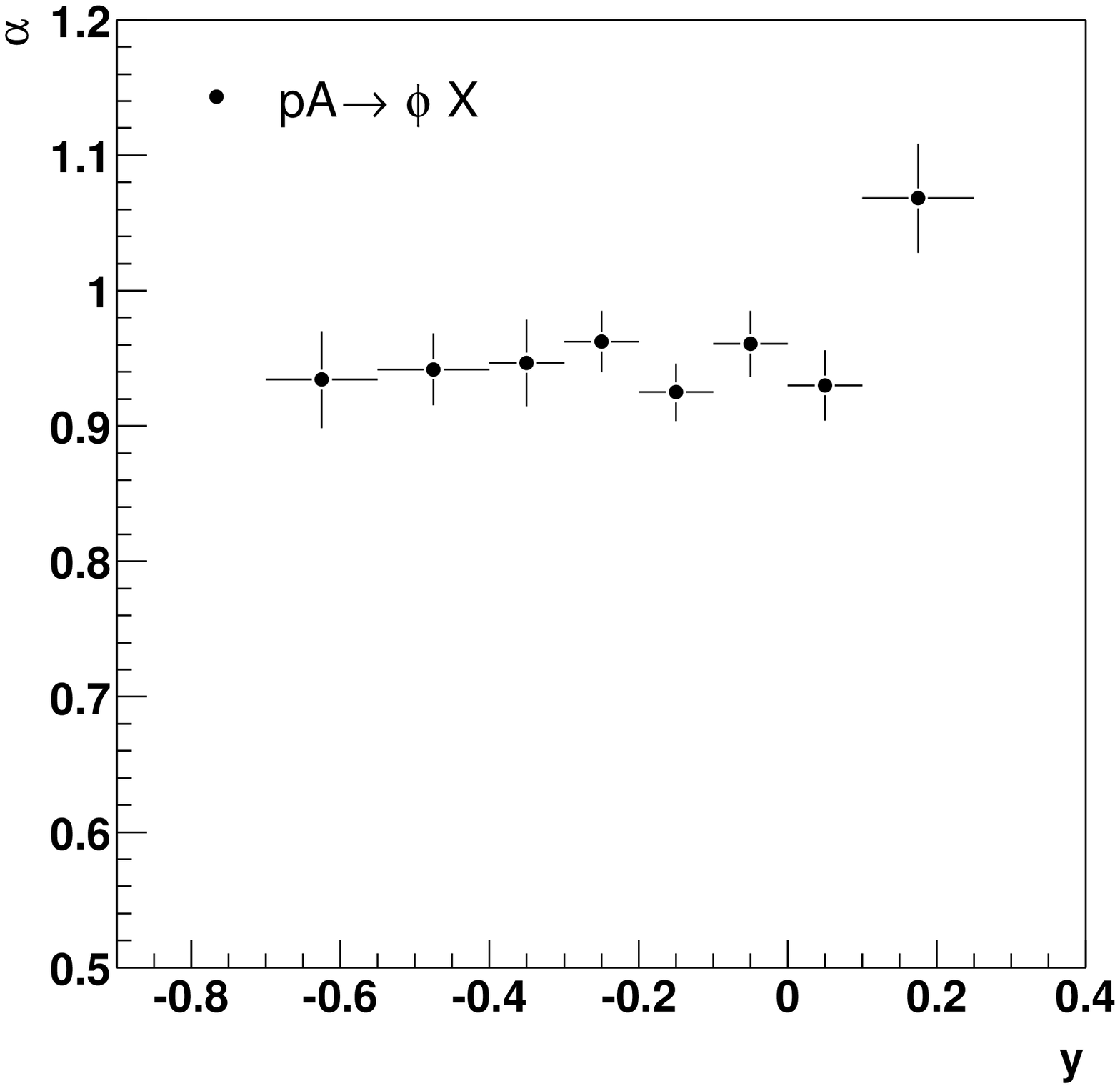}}
  \end{center}
  \caption{Measured values of $\alpha(y)$ for a) 
    \kss/\kssb\ and b) $\phi$ production.}
  \label{fig:alpha_rap}
\end{figure}

The $A$-dependence of the differential cross sections $d\sigma/dy$ and
$d\sigma/dp_T^2$ themselves is also of interest. 
Fig. \ref{fig:alpha_rap} demonstrates that, for the differential
cross sections integrated over $p_T^2$, the exponent $\alpha$ is 
in good approximation independent of rapidity in the range
of the measurement. Note that mainly soft interactions with $p_T^2\le
\unit[1]{GeV^2/c^2}$ contribute to the cross section.

\begin{figure}
  \begin{center}
    \vspace{-7mm}
    \subfigure[]{
      \includegraphics[clip,width=0.4\textwidth]
      {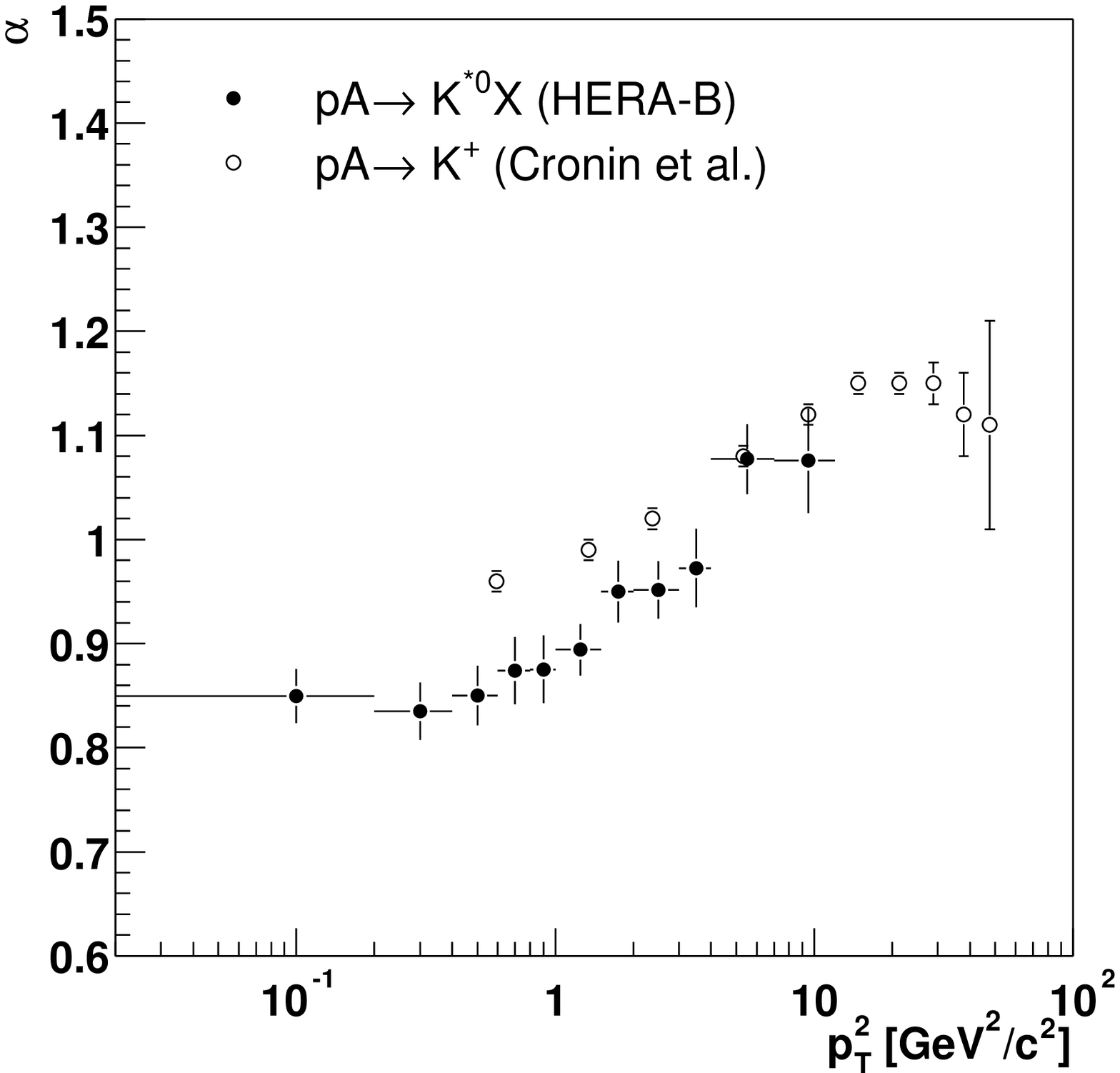}}
    \vspace{-7mm}\\
    \subfigure[]{
      \includegraphics[clip,width=0.4\textwidth]
      {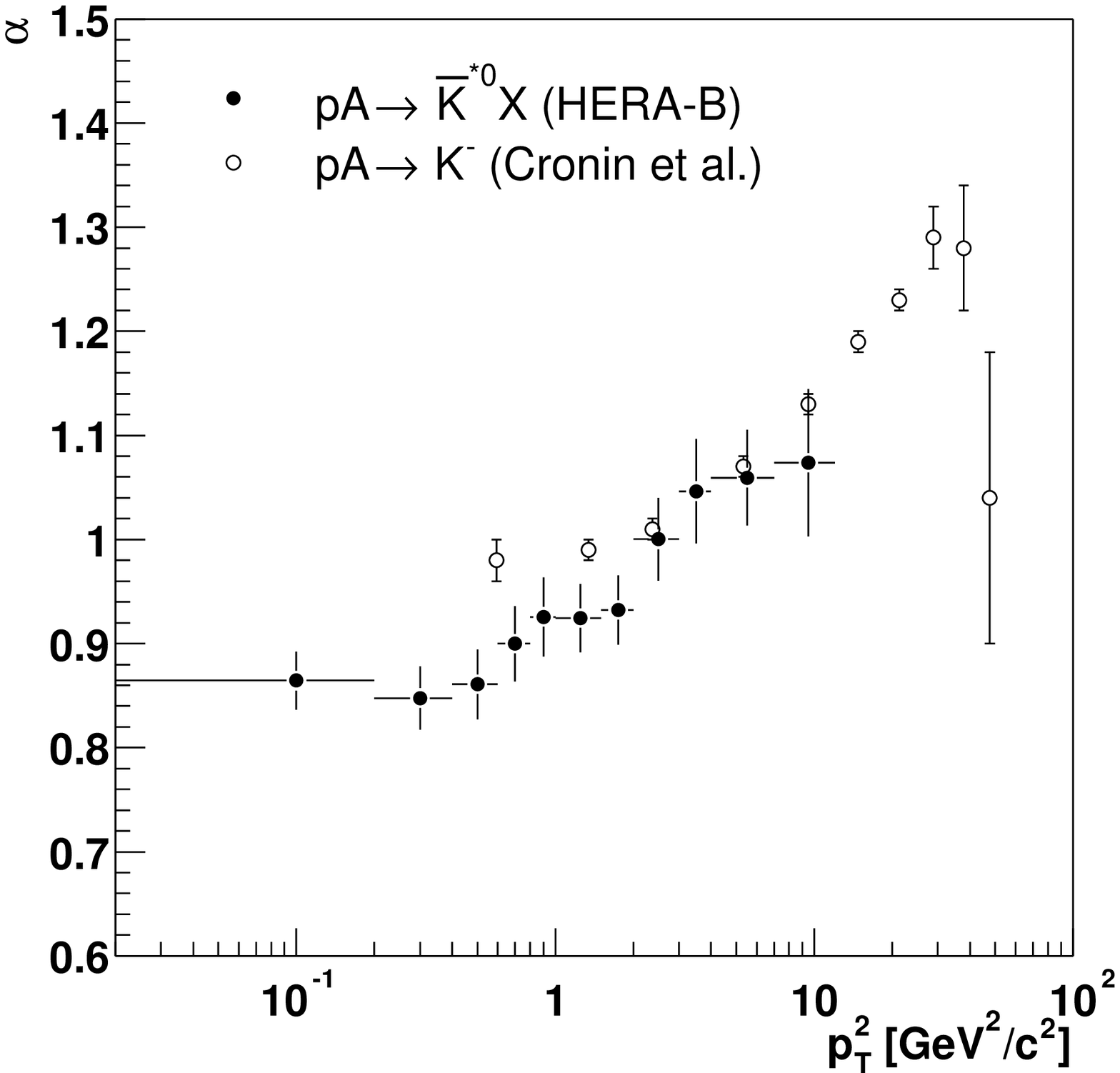}}
    \vspace{-7mm}\\
    \subfigure[]{
      \includegraphics[clip,width=0.4\textwidth]
      {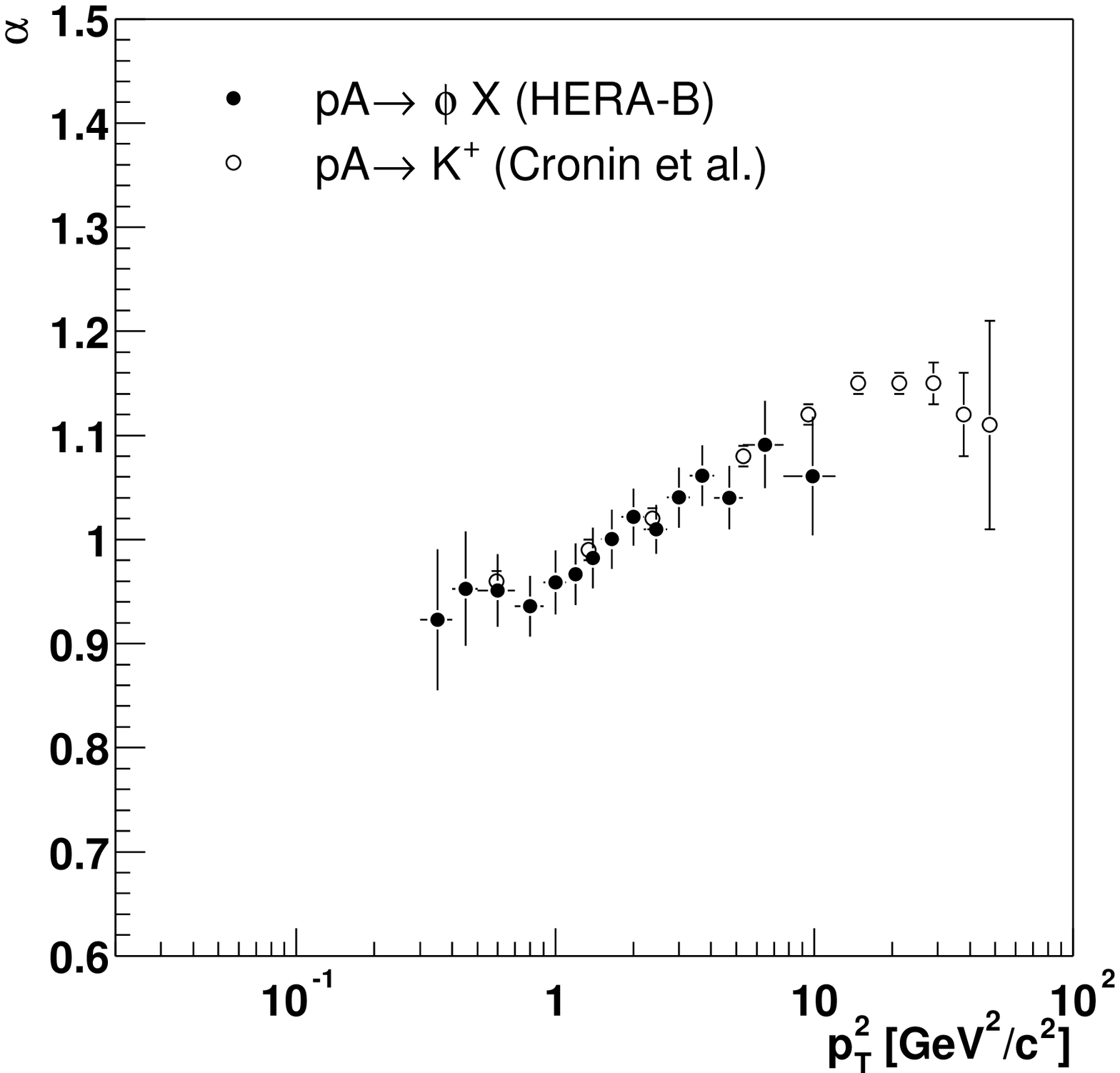}}
  \end{center}
  \caption{Measured values of $\alpha(p_T)$
    for a) \kss, b) \kssb\ and c) $\phi$
    production. The results of Cronin et al. \cite{cro77} on the
    $K^{\pm}$ mesons are superimposed. Numbers of this analysis are
    summarised in Table \ref{tab:CrossSectionSummary2}.}  
  \label{fig:alpha_pt2}
\end{figure}

In contrast, $\alpha$ depends on the transverse momentum.
With increasing $p_T^2$,
$\alpha$ constantly increases, as shown in Fig. \ref{fig:alpha_pt2}.
Cronin et al. \cite{cro77} have observed this
effect for the production of stable particles ($\pi$, $K$, $p$). 
In this analysis we report an observation of the
Cronin effect for the production of vector mesons containing strange
quarks. Note however, that the exponent for \kss\ and \kssb\
production is smaller than that for $K^+$ and $K^-$ mesons. Only at
the highest $p_T$ values do the distributions get close to each other,
while the exponents $\alpha$ for $\phi$ and $K^+$ mesons coincide for
all transverse momenta.  

Qualitatively the observations shown in Fig. \ref{fig:alpha_pt2} are
in agreement with the model of \cite{bzk02} which predicts a decrease
of $\alpha$ with increasing CMS energy; this is indeed observed
for the \kss\ (\kssb) data collected at a beam energy of 920 GeV when
compared to the $K^+$ ($K^-$) results of Cronin et al. measured at 400
GeV. On the other hand $\alpha_\phi$ produced at a beam energy of 920
GeV coincides with that of $K^+$ mesons collected at 400 GeV; this
might indicate a different flavor composition of the initially
scattered partons fragmenting to $\phi$ mesons as compared to \kss\
(\kssb) mesons. 

\section{Conclusion}
The cross sections of \kss, \kssb, and $\phi$ meson production in the
central rapidity region have been measured for the first time in
pC-, pTi-, and pW-interactions. The cross sections show a
power-law dependence $\sigma\propto A^\alpha$ with an exponent of
$\alpha\approx 0.8-1.1$ depending both on the produced vector meson
and the transverse momentum. The exponent $\alpha$ is found to be
larger than the corresponding value for the total inelastic cross
section. The measurements of $\alpha(p_T^2)$ for the $\phi$ vector
meson agree with the values determined by Cronin et al. \cite{cro77}
for $K^+$ mesons. For the \kss/\kssb\ mesons, $\alpha(p_T^2)$ is
systematically smaller. 

\section{Acknowledgements}

We thank B. Kopeliovich for helpful comments and discussions about the
interpretation of our results.
We express our gratitude to the DESY laboratory for the strong support in
setting up and running the HERA-B experiment. We are also indebted to the
DESY accelerator group for the continuous efforts to provide good and
stable beam conditions. 
The HERA-B experiment would not have been possible without the enormous
effort and commitment of our technical and administrative staff. It is a
pleasure to thank all of them. \\

\begin{table*}
  \caption{Compilation of the differential production cross sections
    $d\sigma/dy$ in the accessible phase space of the measurement. The
    errors reflect the statistical and full systematic errors, respectively.} 
  \label{tab:CrossSectionSummary}
  \centering
  \begin{tabular}{c c c c c }
    \hline\noalign{\smallskip}
    & $\Delta y$ & \multicolumn{3}{c}{$\unit[d\sigma/dy]{[mb]}$} \\
    \hline\noalign{\smallskip}
    & & C & Ti & W \\
    \hline\noalign{\smallskip}
    \kss
 & -0.8 -- -0.6 & $12.5\pm 0.4\pm 0.9$ & $38.6\pm 2.2\pm 2.9$ & $128.3\pm 4.6\pm 8.6$ \\
 & -0.6 -- -0.5 & $12.3\pm 0.5\pm 0.9$ & $41.6\pm 2.2\pm 3.2$ & $131.6\pm 5.4\pm 9.0$ \\
 & -0.5 -- -0.4 & $10.8\pm 0.4\pm 0.8$ & $34.0\pm 1.8\pm 2.5$ & $112.9\pm 3.7\pm 7.3$ \\
 & -0.4 -- -0.3 & $12.2\pm 0.3\pm 0.9$ & $39.7\pm 1.7\pm 2.9$ & $120.6\pm 3.4\pm 7.7$ \\
 & -0.3 -- -0.2 & $11.1\pm 0.3\pm 0.8$ & $34.4\pm 1.7\pm 2.4$ & $114.3\pm 3.0\pm 7.1$ \\
 & -0.2 -- -0.1 & $10.6\pm 0.3\pm 0.7$ & $33.5\pm 1.4\pm 2.3$ & $112.8\pm 3.0\pm 7.0$ \\
 & -0.1 -- 0.0 & $10.7\pm 0.3\pm 0.7$ & $34.6\pm 1.4\pm 2.4$ & $115.2\pm 2.9\pm 7.1$ \\
 & 0.0 -- +0.1 & $11.1\pm 0.3\pm 0.8$ & $33.5\pm 1.5\pm 2.4$ & $121.2\pm 3.3\pm 7.6$ \\
 & +0.1 -- +0.2 & $10.1\pm 0.3\pm 0.7$ & $32.4\pm 1.5\pm 2.4$ & $110.3\pm 3.6\pm 7.2$ \\
 & +0.2 -- +0.3 & $10.0\pm 0.4\pm 0.8$ & $37.0\pm 2.3\pm 2.8$ & $109.0\pm 3.9\pm 7.5$ \\

    \hline\noalign{\smallskip}
    \kssb & -0.8 -- -0.6 & $8.5\pm 0.4\pm 0.7$ & $28.7\pm 1.7\pm 2.3$ & $96.8\pm 4.1\pm 7.1$ \\
 & -0.6 -- -0.5 & $9.9\pm 0.5\pm 0.8$ & $30.0\pm 2.1\pm 2.5$ & $114.0\pm 4.9\pm 8.6$ \\
 & -0.5 -- -0.4 & $9.2\pm 0.4\pm 0.7$ & $31.5\pm 1.8\pm 2.5$ & $97.7\pm 4.3\pm 6.9$ \\
 & -0.4 -- -0.3 & $8.8\pm 0.4\pm 0.7$ & $26.6\pm 1.6\pm 2.1$ & $102.9\pm 4.0\pm 7.4$ \\
 & -0.3 -- -0.2 & $9.4\pm 0.3\pm 0.7$ & $31.1\pm 1.5\pm 2.3$ & $102.8\pm 3.4\pm 6.7$ \\
 & -0.2 -- -0.1 & $9.3\pm 0.3\pm 0.7$ & $31.5\pm 1.8\pm 2.4$ & $105.6\pm 3.5\pm 7.0$ \\
 & -0.1 -- 0.0 & $9.5\pm 0.3\pm 0.7$ & $28.4\pm 1.5\pm 2.1$ & $91.7\pm 3.3\pm 6.1$ \\
 & 0.0 -- +0.1 & $8.7\pm 0.3\pm 0.6$ & $26.0\pm 1.3\pm 1.9$ & $88.8\pm 3.4\pm 5.6$ \\
 & +0.1 -- +0.2 & $8.9\pm 0.3\pm 0.7$ & $25.1\pm 1.6\pm 2.0$ & $97.3\pm 3.2\pm 7.2$ \\
 & +0.2 -- +0.3 & $8.1\pm 0.4\pm 0.7$ & $25.9\pm 1.8\pm 2.2$ & $92.3\pm 4.0\pm 7.1$ \\

    \hline\noalign{\smallskip}
    $\phi$
 & -0.70 -- -0.55 & $1.77\pm 0.10\pm 0.18$ & $7.2\pm 0.6\pm 0.8$ & $23.1\pm 0.9\pm 2.1$ \\
 & -0.55 -- -0.40 & $1.81\pm 0.06\pm 0.17$ & $7.0\pm 0.4\pm 0.7$ & $24.0\pm 0.6\pm 2.1$ \\
 & -0.40 -- -0.30 & $1.80\pm 0.06\pm 0.21$ & $7.2\pm 0.3\pm 0.7$ & $24.7\pm 0.7\pm 2.2$ \\
 & -0.30 -- -0.20 & $1.60\pm 0.05\pm 0.14$ & $6.8\pm 0.3\pm 0.7$ & $23.0\pm 0.5\pm 2.0$ \\
 & -0.20 -- -0.10 & $1.71\pm 0.05\pm 0.14$ & $6.7\pm 0.3\pm 0.6$ & $21.7\pm 0.5\pm 1.9$ \\
 & -0.10 -- 0.00 & $1.64\pm 0.05\pm 0.14$ & $6.4\pm 0.3\pm 0.6$ & $22.7\pm 0.6\pm 2.0$ \\
 & 0.00 -- +0.10 & $1.84\pm 0.07\pm 0.16$ & $7.3\pm 0.4\pm 0.7$ & $23.6\pm 0.8\pm 2.1$ \\
 & +0.10 -- +0.25 & $1.45\pm 0.10\pm 0.14$ & $6.1\pm 0.5\pm 0.6$ & $26.5\pm 1.1\pm 2.5$ \\

    \hline\noalign{\smallskip}
  \end{tabular}
\end{table*}

\begin{table*}
  \caption{Compilation of the differential production cross sections
    $d\sigma/dp_T^2$ in the accessible phase space of the
    measurement. The errors reflect the statistical and full
    systematic errors, respectively. The right column lists the
    results of the $A^\alpha$ fits shown in
    Figs. \ref{fig:alpha_pt2}. The errors given are the combined
    statistical and target-material dependent systematic errors,
    added in quadrature.} 
  \label{tab:CrossSectionSummary2}
  \centering
  \begin{tabular}{c c c c c | c}
    \hline\noalign{\smallskip}
    & \unit[$\Delta(p_T^2)]{[GeV^2/c^2]}$ &
    \multicolumn{4}{c}{$\unit[d\sigma/dp_T^2]{[mb/(GeV/c)^2]}$} \\
    \hline\noalign{\smallskip}
    & & C & Ti & W & $\alpha$\\
    \hline\noalign{\smallskip}
    \kss
 & 0.0 -- 0.2 & $22.2\pm 0.5\pm 1.5$ & $69.4\pm 4.3\pm 4.9$ & $224.3\pm 5.8\pm 14.0$& $0.85\pm 0.03$ \\
 & 0.2 -- 0.4 & $12.2\pm 0.4\pm 0.8$ & $38.9\pm 1.6\pm 2.7$ & $119.4\pm 4.2\pm 7.4$& $0.84\pm 0.03$ \\
 & 0.4 -- 0.6 & $7.4\pm 0.2\pm 0.5$ & $25.5\pm 1.3\pm 1.8$ & $76.0\pm 2.7\pm 4.9$& $0.85\pm 0.03$ \\
 & 0.6 -- 0.8 & $4.7\pm 0.2\pm 0.3$ & $14.5\pm 0.8\pm 1.1$ & $50.5\pm 2.0\pm 3.3$& $0.87\pm 0.03$ \\
 & 0.8 -- 1.0 & $3.5\pm 0.1\pm 0.3$ & $10.6\pm 0.7\pm 0.8$ & $37.9\pm 1.3\pm 2.5$& $0.88\pm 0.03$ \\
 & 1.0 -- 1.5 & $1.92\pm 0.04\pm 0.13$ & $6.4\pm 0.2\pm 0.4$ & $21.8\pm 0.5\pm 1.3$& $0.89\pm 0.02$ \\
 & 1.5 -- 2.0 & $0.91\pm 0.03\pm 0.07$ & $3.1\pm 0.2\pm 0.2$ & $11.9\pm 0.3\pm 0.8$& $0.95\pm 0.03$ \\
 & 2.0 -- 3.0 & $0.37\pm 0.01\pm 0.03$ & $1.45\pm 0.06\pm 0.11$ & $5.0\pm 0.1\pm 0.3$& $0.95\pm 0.03$ \\
 & 3.0 -- 4.0 & $0.157\pm 0.006\pm 0.013$ & $0.56\pm 0.04\pm 0.05$ & $2.20\pm 0.08\pm 0.16$& $0.97\pm 0.04$ \\
 & 4.0 -- 7.0 & $0.035\pm 0.001\pm 0.003$ & $0.142\pm 0.008\pm 0.011$ & $0.64\pm 0.02\pm 0.04$& $1.08\pm 0.03$ \\
 & 7.0 -- 12.0 & $0.0044\pm 0.0003\pm 0.0004$ & $0.021\pm 0.002\pm 0.002$ & $0.084\pm 0.004\pm 0.007$& $1.08\pm 0.05$ \\

    \hline\noalign{\smallskip}
    \kssb & 0.0 -- 0.2 & $19.3\pm 0.5\pm 1.4$ & $59.2\pm 3.4\pm 4.3$ & $201.8\pm 5.5\pm 13.0$& $0.86\pm 0.03$ \\
 & 0.2 -- 0.4 & $10.1\pm 0.3\pm 0.7$ & $31.0\pm 1.5\pm 2.2$ & $101.3\pm 3.7\pm 6.5$& $0.85\pm 0.03$ \\
 & 0.4 -- 0.6 & $6.2\pm 0.2\pm 0.4$ & $19.1\pm 1.1\pm 1.4$ & $65.0\pm 2.9\pm 4.3$& $0.86\pm 0.03$ \\
 & 0.6 -- 0.8 & $4.0\pm 0.2\pm 0.3$ & $13.9\pm 0.9\pm 1.1$ & $46.2\pm 2.3\pm 3.2$& $0.90\pm 0.04$ \\
 & 0.8 -- 1.0 & $2.8\pm 0.1\pm 0.2$ & $10.1\pm 0.7\pm 0.8$ & $34.5\pm 1.7\pm 2.5$& $0.93\pm 0.04$ \\
 & 1.0 -- 1.5 & $1.55\pm 0.04\pm 0.12$ & $5.1\pm 0.3\pm 0.4$ & $19.0\pm 0.7\pm 1.3$& $0.92\pm 0.03$ \\
 & 1.5 -- 2.0 & $0.68\pm 0.02\pm 0.05$ & $2.6\pm 0.1\pm 0.2$ & $8.7\pm 0.3\pm 0.6$& $0.93\pm 0.03$ \\
 & 2.0 -- 3.0 & $0.264\pm 0.010\pm 0.022$ & $0.88\pm 0.05\pm 0.07$ & $3.9\pm 0.1\pm 0.3$& $1.00\pm 0.04$ \\
 & 3.0 -- 4.0 & $0.102\pm 0.006\pm 0.011$ & $0.45\pm 0.03\pm 0.04$ & $1.78\pm 0.08\pm 0.16$& $1.05\pm 0.05$ \\
 & 4.0 -- 7.0 & $0.024\pm 0.001\pm 0.002$ & $0.097\pm 0.008\pm 0.009$ & $0.42\pm 0.02\pm 0.03$& $1.06\pm 0.05$ \\
 & 7.0 -- 12.0 & $0.0018\pm 0.0002\pm 0.0002$ & $0.010\pm 0.002\pm 0.001$ & $0.035\pm 0.003\pm 0.003$& $1.07\pm 0.07$ \\

    \hline\noalign{\smallskip}
    $\phi$
 & 0.3 -- 0.4 & $2.8\pm 0.3\pm 0.2$ & $9.4\pm 1.7\pm 1.2$ & $34.9\pm 4.1\pm 3.8$& $0.92\pm 0.07$ \\
 & 0.4 -- 0.5 & $2.3\pm 0.2\pm 0.3$ & $9.2\pm 1.3\pm 1.1$ & $31.2\pm 2.5\pm 3.1$& $0.95\pm 0.06$ \\
 & 0.5 -- 0.7 & $1.70\pm 0.09\pm 0.16$ & $7.0\pm 0.6\pm 0.7$ & $23.0\pm 1.1\pm 2.1$& $0.95\pm 0.03$ \\
 & 0.7 -- 0.9 & $1.09\pm 0.05\pm 0.09$ & $4.5\pm 0.3\pm 0.5$ & $14.2\pm 0.6\pm 1.3$& $0.94\pm 0.03$ \\
 & 0.9 -- 1.1 & $0.73\pm 0.03\pm 0.07$ & $2.8\pm 0.2\pm 0.3$ & $10.1\pm 0.4\pm 0.9$& $0.96\pm 0.03$ \\
 & 1.1 -- 1.3 & $0.46\pm 0.02\pm 0.04$ & $2.0\pm 0.1\pm 0.2$ & $6.6\pm 0.2\pm 0.6$& $0.97\pm 0.03$ \\
 & 1.3 -- 1.5 & $0.35\pm 0.02\pm 0.03$ & $1.57\pm 0.09\pm 0.16$ & $5.2\pm 0.2\pm 0.5$& $0.98\pm 0.03$ \\
 & 1.5 -- 1.8 & $0.266\pm 0.010\pm 0.024$ & $1.02\pm 0.06\pm 0.10$ & $4.0\pm 0.1\pm 0.4$& $1.00\pm 0.03$ \\
 & 1.8 -- 2.2 & $0.163\pm 0.006\pm 0.015$ & $0.62\pm 0.03\pm 0.06$ & $2.59\pm 0.07\pm 0.23$& $1.02\pm 0.03$ \\
 & 2.2 -- 2.7 & $0.098\pm 0.001\pm 0.009$ & $0.43\pm 0.02\pm 0.04$ & $1.57\pm 0.04\pm 0.14$& $1.01\pm 0.02$ \\
 & 2.7 -- 3.3 & $0.058\pm 0.002\pm 0.005$ & $0.24\pm 0.01\pm 0.02$ & $0.99\pm 0.03\pm 0.09$& $1.04\pm 0.03$ \\
 & 3.3 -- 4.1 & $0.031\pm 0.001\pm 0.003$ & $0.145\pm 0.009\pm 0.014$ & $0.57\pm 0.02\pm 0.05$& $1.06\pm 0.03$ \\
 & 4.1 -- 5.3 & $0.0157\pm 0.0008\pm 0.0014$ & $0.065\pm 0.005\pm 0.007$ & $0.268\pm 0.010\pm 0.024$& $1.04\pm 0.03$ \\
 & 5.3 -- 7.6 & $0.0049\pm 0.0003\pm 0.0005$ & $0.021\pm 0.002\pm 0.002$ & $0.095\pm 0.004\pm 0.009$& $1.09\pm 0.04$ \\
 & 7.6 -- 12.0 & $0.0010\pm 0.0001\pm 0.0001$ & $0.0063\pm 0.0010\pm 0.0009$ & $0.0183\pm 0.0008\pm 0.0020$& $1.06\pm 0.06$ \\

    \hline\noalign{\smallskip}
  \end{tabular}
\end{table*}

\end{document}